%% file: main.tex
\documentclass{article}

\pdfoutput=1

\usepackage{arxiv}

\usepackage[numbers]{natbib}

\usepackage[utf8]{inputenc} 
\usepackage[T1]{fontenc}    
\usepackage{hyperref}       
\usepackage{url}            
\usepackage{booktabs}       
\usepackage{amsfonts}       
\usepackage{nicefrac}       
\usepackage{microtype}      
\usepackage{lipsum}		
\usepackage{graphicx}
\usepackage{natbib}
\usepackage{doi}
\usepackage{bm}

\usepackage{amsmath}
\usepackage{amssymb}

 \usepackage{enumitem}

\newtheorem{remark}{Remark}%
\newtheorem{principle}{Principle}%
\newtheorem{rules}{Rule}%

\newcommand{\bx}{\bm{x}}

\newcommand{\by}{\bm{y}}
\newcommand{\bz}{\bm{z}}

\newcommand{\sX}{\mathcal{X}}

\newcommand{\coo }{\ensuremath{\mathrm{CO_2} }}

\title{ {Towards Carbon-Free Electricity:} \\ A Flow-Based Framework for Power Grid Carbon Accounting and Decarbonization}

\date{} 					

\author{ {Xin Chen}\thanks{corresponding author.} \\
	Department of Electrical and Computer Engineering\\
	Texas A\&M University\\
	College Station, TX, 77843, USA. \\
	Email: \texttt{xin\_chen@tamu.edu} \\
	\And
	{Hungpo Chao} \\
	Energy Trading Analytics, LLC\\
	Stanford University (ret.)\\
	Palo Alto, CA, USA. \\
	\texttt{ } \\
 	\And
	{Wenbo Shi} \\
	Singularity Energy Inc.\\
	Somerville, MA, 02143, USA. \\
  	\And
	{Na Li} \\
	School of Engineering and Applied Sciences\\
	Harvard University\\
	Cambridge, MA, 02134, USA.\\
}

\renewcommand{\headeright}{ }
\renewcommand{\shorttitle}{ }

\hypersetup{
pdftitle={A template for the arxiv style},
pdfsubject={q-bio.NC, q-bio.QM},
pdfauthor={David S.~Hippocampus, Elias D.~Striatum},
pdfkeywords={First keyword, Second keyword, More},
}

\begin{document}
\maketitle

\begin{abstract}
 This paper introduces a comprehensive framework aimed at advancing research and policy development in the realm of decarbonization within electric power systems. The framework focuses on three key aspects: carbon accounting, carbon-aware decision-making, and carbon-electricity market design. It addresses existing problems, methods, and proposes solutions.  
In contrast to traditional pool-based emissions models, our framework proposes a novel flow-based emissions model. This model incorporates the underlying physical power grid and power flows, allowing for accurate carbon accounting at both temporal and spatial scales. This, in turn, facilitates informed decision-making to achieve grid decarbonization goals.  The framework is built on a flow-based accounting methodology and utilizes the carbon-aware optimal power flow (C-OPF) technique as a theoretical foundation for decarbonization decision-making. Additionally, the paper explores the potential design of carbon-electricity markets and pricing mechanisms to incentivize decentralized decarbonization actions. The critical issues of data availability, infrastructure development, and considerations of fairness and equity are also discussed.  
This paper seeks to advance scholarly understanding and foster progress toward achieving sustainable and carbon-free electric power systems.

\end{abstract}


\newpage
\tableofcontents

\newpage
\input{Introduction}

\input{Account}

\input{Decision}

\input{Design}
\input{Conclusion}

\bibliographystyle{unsrt}
\bibliography{references}  






\end{document}

%% file: Introduction.tex
\section{Introduction}


Human-induced climate change and global warming, primarily caused by the increasing levels of carbon dioxide (\coo) and other greenhouse gases (GHG) in the atmosphere, are posing serious threats to the well-being of billions of people around the world and  leading to widespread and irreversible disruption in nature \cite{portner2022climate}.  
Decarbonization is an urgent priority  in order to mitigate these impacts and ensure a sustainable future for the environment and human society. In 2022, the 
the world emitted around 57.8 billion tonnes of greenhouse gases (measured in carbon dioxide equivalents), 
35.6\% of which came from the energy systems and were primarily caused by the combustion of fossil fuels for electricity production\cite{carbonclock}. 
The supply and utilization of low-carbon electricity  are necessary to achieving substantial  reductions in carbon emissions. Hence, the electric power sector, involving the generation, transmission,  distribution, and consumption of electricity, 
plays a crucial role in the mission of decarbonization.

\begin{figure} %
\centering
\includegraphics[scale=0.53]{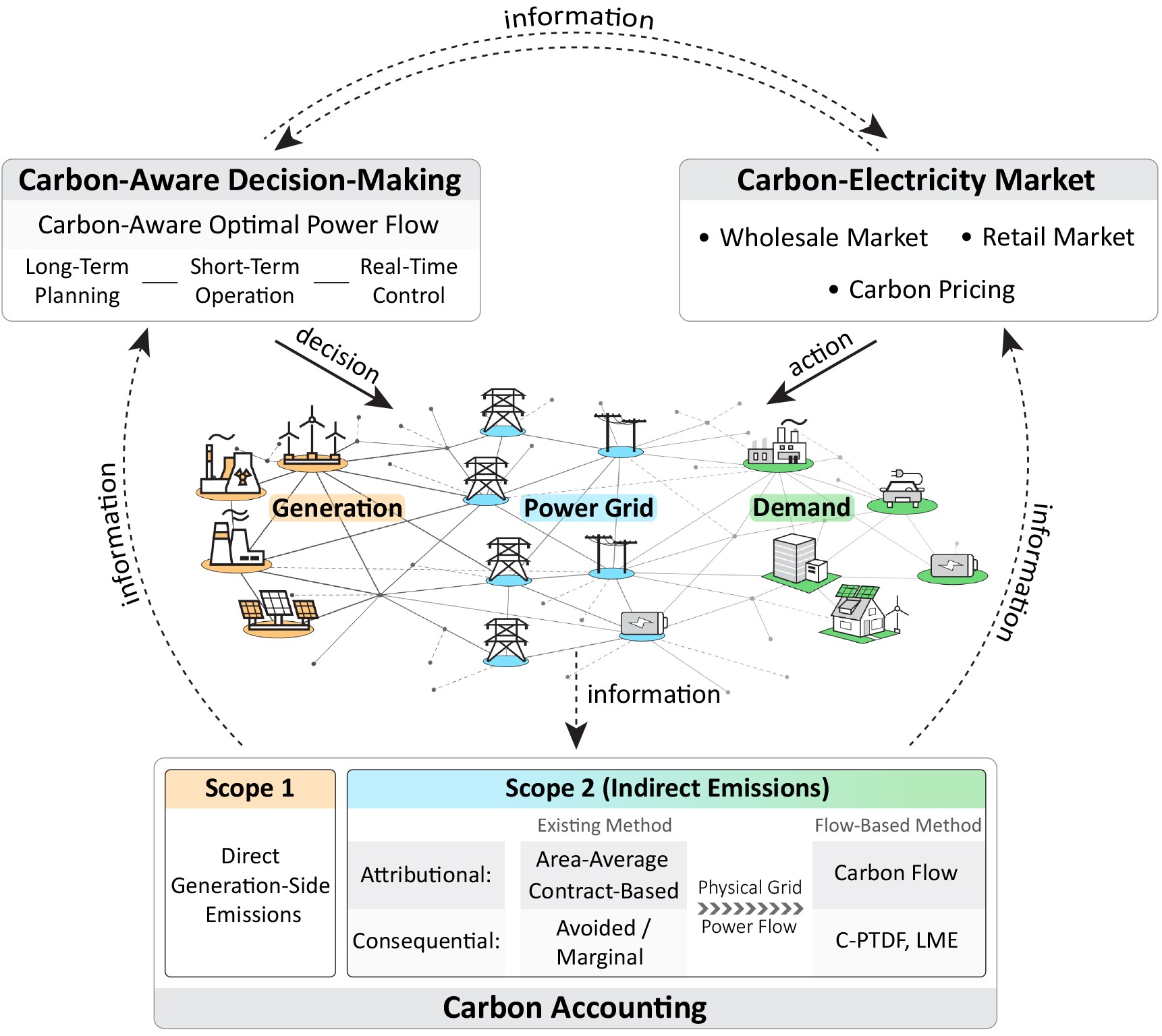}
\caption{The proposed carbon research and development framework for carbon-electricity nexus. (C-PTDF: Carbon-Power Transfer Distribution Factor, LME: Locational Marginal Emissions.) }
\label{fig:research}
\end{figure}

To decarbonize electric power systems, a variety of technologies have been deployed, such as increasing penetration of renewable energy sources, applying carbon capture, utilization, and storage (CCUS) facilities, improving generation efficiency, etc. 
In addition to these technical approaches on the generation side, it is essential to harness the tremendous resources available from a wide range of end-users on the \emph{demand} side for achieving effective decarbonization. 
In power systems, electricity generation and consumption must be balanced at all times and in all places of the grid to ensure stable and reliable system operation. Therefore, although almost all carbon emissions occur on the generation side, electricity consumption is the inherent drive of power generation that results in carbon emissions. In recent decades, the demand side has witnessed a rapid proliferation of distributed energy resources (DERs), such as smart buildings, energy storage, electric vehicles, responsive loads, solar panels, etc. The coordination of massive DERs can provide enormous power flexibility to facilitate low-carbon power system operation. Moreover, an ever-expanding group of energy consumers, comprising  corporations, municipalities, institutions, and individuals, are  
     dedicated to the goal of achieving carbon-free electricity, such as Google's 24/7 carbon-free energy program \cite{google247program},  Microsoft's  carbon-free
electricity globally project \cite{microsoft}, 
the U.S. federal government's Executive Order 14057\footnote{Executive Order 14057 - Catalyzing Clean Energy Industries and Jobs Through Federal Sustainability,
December 8, 2021.}. It presents unprecedented opportunities to leverage investments, advanced technologies, abundant resources, and collaborative efforts from vast end-users to decarbonize the electric power sector.

The integration of carbon emissions analytics into the planning, operation, control, and market of electricity systems is referred to as ``Carbon-Electricity Nexus''. This field is emerging and presents significant challenges owing to the complexity of large-scale engineering systems, physical laws and constraints, policy considerations, economic factors, market dynamics, carbon accounting principles, the involvement of multi-stakeholders, and more. To this end, this paper aims to establish an overarching framework to guide the research and development in carbon-electricity nexus. Specifically, 
as shown in Figure \ref{fig:research},
our framework breaks down the intricate subject of carbon-electricity nexus into three fundamental tasks:

\setlist{leftmargin=6mm}
\begin{itemize} 
    \item [1)] \emph{Carbon Accounting}: 
    This task involves quantifying the carbon emissions from electricity generation, as well as attributing these emissions to end users to clarify the emission responsibility
associated with electricity consumption. 
    
  
    \item [2)] \emph{Decarbonization Decision-Making}: The development of effective 
     decision-making schemes is necessitated for power system  planning, operation, control, and energy management to optimize the decarbonization process. 
    \item [3)]  \emph{Carbon-Electricity Market Design}: 
    It is crucial to integrate
    the goal of decarbonization into electricity markets, to 
account for the externality of carbon emissions and foster economic incentives
for both suppliers and consumers to pursue low-carbon products and services.

    

\end{itemize}

The three fundamental tasks above are interconnected. Essentially, 
 \emph{carbon accounting} answers the question that what carbon emission amounts are for each entity;  \emph{decarbonization decision-making} determines what the right actions are to reduce these emissions; and \emph{carbon-electricity market design}  identifies economic mechanisms to incentivize various entities to undertake such actions for decarbonization. 
 There have been a number of studies for each of these three tasks. However, most existing approaches and mechanisms suffer from two  critical limitations: 1) 
\emph{lack of spatial and temporal granularity}, and 2) \emph{disconnect with physical power systems}. 
As a result, these approaches may fail to deliver trustworthy carbon emission information, 
potentially leading to decarbonization initiatives and market instruments that do not yield substantial carbon emissions reductions   despite heavy investments\cite{brander2018creative,bjorn2022rec}. 
Hence, our proposed framework for carbon-electricity nexus emphasizes the \emph{alignment of carbon accounting, decision-making, and market strategies with the physical power systems and underlying power flow}. This feature is termed ``\emph{flow-based}'',  distinguishing it from conventional ``pool-based" carbon accounting and decarbonization schemes. 
We elaborate on the proposed flow-based framework in the following sections:
 \begin{itemize}
     \item [1)] For carbon accounting, Section \ref{sec:accounting} provides a comprehensive overview of existing methods with in-depth discussions of their advantages and limitations, and presents the flow-based carbon accounting methodology.
     \item [2)] For carbon-aware decision-making, Section \ref{sec:decision} presents the carbon-aware optimal power flow (C-OPF) method as the theoretical foundation, and frames the critical decarbonization decision problems from the perspectives of power 
generators, grid operators, and end-users across different time scales.
\item [3)] For carbon-electricity market design, Section \ref{sec:market} delves into the fundamentals of wholesale and retail electricity markets and explores potential ways to integrate carbon emissions into these markets and carbon-electricity pricing. 
 \end{itemize}

  Lastly, Section \ref{sec:conclusion} discusses the critical challenges and future vision 
  in the implementation of 
the flow-based framework and other effective decarbonization schemes. 
  This proposed framework for
  carbon-electricity nexus can serve as a reference for future research, education programs, policy development, and business initiatives concerning carbon emissions in the electric power sector. It provides guidance and insights for developing effective decarbonization strategies, contributing to the ultimate goal of achieving carbon-free electricity and a sustainable future.

%% file: Account.tex
\section{Carbon Accounting }\label{sec:accounting}

Carbon (emissions) accounting is the process of measuring and quantifying the amount of carbon emissions associated with specific activities, which lays the quantitative foundation for carbon-electricity nexus. In addition to accurately measuring the carbon emissions produced by power generators (known as generation-side carbon accounting), it is essential to properly and fairly attribute the generation-side emissions to end-users according to their electricity consumption, which is referred to as \emph{demand-side carbon accounting}. On the one hand, 
electricity consumption intrinsically creates the need for power generations, resulting in carbon emissions from the combustion of fossil fuels. On the other hand, demand-side carbon accounting is a crucial prerequisite for engaging vast users in decarbonization initiatives, regulatory policies, and carbon-electricity markets. Accordingly,
the GHG Protocol \cite{world2004greenhouse,world2014scope2} defines two categories of GHG emissions,  Scope 1 and Scope 2,  to distinguish direct generation-side emissions and indirect (attributed) demand-side emissions for the purposes of carbon accounting and reporting. 
 Specifically, in the electric power sector, {power plant owners} report Scope 1 emissions that result from the generation of electricity, heat, steam, etc. {Power grid owners} and {utility companies} report Scope 2 emissions 
 associated with  the transmission and distribution  loss of the power grids that they own or control. {End-users} report Scope 2 emissions associated with purchased electricity, steam, heating,  cooling, etc. 
 This paper focuses on the carbon accounting for Scope 2 emissions on the demand side.

Furthermore, there are two primary types of carbon accounting frameworks: \emph{attributional} and \emph{consequential}.  The \emph{attributional} carbon accounting is designed to allocate direct generation-side emissions to end-users, such as  the  \emph{location-based} method and \emph{market-based} method introduced below.
In contrast, the \textit{consequential} carbon accounting is developed to identify and quantify the emissions impact of a specific decision or project, such as the \emph{marginal emission} approaches and \emph{avoided emission} accounting. 
Nevertheless, these two carbon accounting frameworks are closely related despite for different purposes, and we explain their underlying relations in Section \ref{sec:f-consequential}.



In this section, we first introduce three guiding principles to enable trustworthy and informative carbon accounting. We then review existing widely-used carbon accounting methods, along with discussions on their advantages and disadvantages. Next, we propose a generic flow-based carbon accounting approach that aligns with the  physical power systems and underlying power flows. 
At last, we conclude by presenting the critical issues regarding carbon accounting in the electric power sector.


\subsection{Fundamental Carbon Accounting Principles}



Demand-side (Scope 2) carbon accounting is challenging and complex, as the electricity systems are large-scale interconnected socio-technical systems with various generators, consumers, and stakeholders. Moreover, demand-side carbon emissions are virtually allocated quantities that cannot be measured physically. As a result, numerous carbon accounting methods based on diverse mechanisms have been proposed to quantify demand-side emissions, often yielding significantly varying carbon emission figures. 
Trustworthy and informative carbon accounting is necessitated because 1) carbon accounting 
quantifies the emissions responsibility, which associates with real-world social and economic benefits or penalties; 2) carbon accounting informs decarbonization decision-making, and unfaithful accounting results are less informative or may even mislead investments, projects, market design, and policy development. 
Hence, this paper suggests the following three guiding principles for carbon accounting.

\begin{principle}\label{prin:comm}
\textbf{\emph{The total demand-side emissions should be equal to the total generation-side emissions.}}
\end{principle}
While this principle may appear trivial, it underscores a crucial property that the demand-side emissions are not artificial attributes of electricity but should add up and be consistent with the physical generation-side emissions. Therefore, carbon accounting needs to prevent double accounting and the instances when multiple entities claim credits for the same emission reduction.



\begin{principle}\label{prin:physical}
\textbf{\emph{Carbon accounting should sufficiently align with the physical power system operation.}}
\end{principle}
Power grids are physical engineered systems with specific network topology and electric components. The delivery of electricity through a power grid, namely \emph{power flow} \cite{giannakis2013monitoring}, directly connects electricity consumption and generation as well as the resultant carbon emissions. Additionally, electric power flow is not arbitrary but follows certain physics laws
(known as {power flow equations or Kirchhoff's laws} \cite{grainger1999power}) and network constraints (such as transmission capacity limits and voltage constraints). Hence, it is important to align carbon accounting with the physical power system operation and underlying power flow, while the alignment can be adaptive to various use cases and different timescales.




\begin{principle}\label{prin:resolution}
\textbf{\emph{Carbon accounting should have sufficient temporal and spatial resolutions.}}
\end{principle}
The generation fuel mix in power grids is   
constantly changing over time, leading to dynamic carbon emission footprints with daily and seasonal patterns. 
Recent study\cite{miller2022hourly} shows that the use of annual-average carbon accounting can over- or under-estimate carbon emissions up to
35\%, in comparison with hourly accounting. Furthermore, the carbon emission intensity  also 
 varies geographically, as generators with different fuel types are distributed in different locations. To facilitate effective decarbonization policies and decision-making, carbon accounting schemes require sufficient temporal and spatial granularity to capture the variations, patterns, and  trends  in emissions.

\subsection{Current Commonly-Used Accounting Methods}



As introduced in the GHG Protocol \cite{world2004greenhouse}, existing \emph{attributional} carbon accounting schemes include the
\emph{location-based} method and \emph{market-based} methods, while \emph{consequential} accounting involves
 the \emph{marginal emission} methods and \emph{avoided emissions}. These methods are discussed below and summarized in Table \ref{tab:existingmethod}.

\begin{table*}[ht]
\centering
 \caption{Summary of existing carbon accounting methods.}
\label{tab:existingmethod}
\small
\renewcommand{\arraystretch}{1.7}
\begin{tabular}{lccc}
\hline
& \multicolumn{2}{c}{Attributional Accounting} & Consequential Accounting\\ \hline
Methods      & Location-based   & Market-based  (REC, PPA, 24/7 matching) & Avoided  \\
Features     & {\begin{tabular}[c]{@{}c@{}} \textbf{Physical} allocation based on \vspace{-8pt}\\ the generation mix in the area\end{tabular}} &\begin{tabular}[c]{@{}c@{}}\textbf{Contractual} allocation based on \vspace{-8pt}\\ power purchases and energy attribute claims\end{tabular} & \begin{tabular}[c]{@{}c@{}}\textbf{Counter-factual} evaluation in \vspace{-8pt}\\  comparison with a baseline\end{tabular} \\
Calculation  &{\begin{tabular}[c]{@{}c@{}} 
Grid average emission factors \vspace{-8pt}\\ (AEFs)\end{tabular} }                 & \begin{tabular}[c]{@{}c@{}}Utility-/supplier-specific \vspace{-8pt}\\ emission factors\end{tabular}      &\begin{tabular}[c]{@{}c@{}}Grid marginal/consequential emission  \vspace{-8pt}\\ factors (MEFs/CEFs)\end{tabular}              \\
 Data Sources & EPA eGRID, EIA, IEA   & Utilities, EEI   & \begin{tabular}[c]{@{}c@{}}EPA eGRID, AVERT, ISOs, \vspace{-8pt}\\ NREL Cambium, some third parties\end{tabular}                         \\ \hline
\end{tabular}
\end{table*}


\subsubsection{Location-Based Carbon Accounting}\label{sec:location}

The location-based carbon accounting method\cite{world2004greenhouse} defines \textbf{Average Emission Factor} (AEF) of an area grid over a certain time period, to count for demand-side carbon emissions. AEF is calculated by dividing the total carbon emissions by the total electricity generation, which describes how much carbon emissions are produced on average to generate 1 unit of electricity. Then, the carbon emissions attributed to an electricity consumer entity is computed by multiplying the AEF by its electricity consumption.  
The current practice of location-based carbon accounting is on an \emph{annual} basis and averaged across 
 large balancing areas  or electricity market zones, such as the  California ISO (CAISO)  region. 
Other jurisdictional boundaries, such as  North American Electric Reliability Corporation (NERC) regions, eGRID regions, and states, are also used to define AEF in the U.S. There are several available data sources of the AEF data, such as EPA eGRID\cite{eGRID} and EIA datasets\cite{eia}.  Some ISOs (e.g., ISO New England\cite{isone}, CAISO\cite{CAISO}) and companies (e.g., the Open Grid Emissions datasets from Singularity Energ\cite{singu}) also provide the AEF datasets.

 \noindent
\textbf{Discussions.} There are 
two caveats regarding the use of AEF for carbon accounting: 1) It does not consider the physical power grids and power flow delivery, and lacks temporal and spatial granularity when calculated as the annual average across a wide balancing area (violation of Principles \ref{prin:physical} and \ref{prin:resolution}).  2) A power grid generally has power imports and exports with its adjacent grids, and the interchanged power can constitute a significant portion of the overall electricity supply.
    Whether to factor in power imports or not, known as \emph{consumption-based} \cite{de2019tracking} and \emph{production-based}, can substantially affect the calculation of AEF and accounting results. 

\subsubsection{Market-Based Carbon Accounting} \label{sec:marketbase}

The market-based methods account for carbon emissions associated with  electricity that users have contractually purchased. In
contrast to location-based methods, market-based carbon accounting methods reflect the \emph{contractual} information or claims flow, 
and they
derive emission factors from contractual instruments\cite{world2014scope2}, 
regardless of the actual electricity delivery. Commonly-used contractual instruments 
include
energy attribute certificates (e.g., Renewable Energy Certificate (REC)\cite{lau2008bottom,jones2015legal}, Guarantee of Origin (GO)), direct contracts (e.g., Power Purchase Agreement (PPA)\cite{kansal2018introduction,tang2019classification} and its variants), supplier-specific emission rates, etc. For example, if a company procures an unbundled REC or has a 1MWh electricity purchase contract with a renewable energy supplier, market-based accounting allows the company to report zero carbon emissions (or use a zero emission factor) for its 1MWh electricity consumption. 
 As a consequence, location-based methods and market-based methods may yield significantly different carbon accounting results for the same end-user. Hence, 
the GHG Protocol \cite{world2014scope2} requires that  companies engaged in   contractual instruments should report Scope 2 emissions 
in both location-based and market-based methods, 
a practice known as \emph{dual reporting}.

Under market-based carbon accounting, \emph{voluntary procurement of renewable energy}  has emerged as a popular strategy employed by large electricity consumers to achieve their  decarbonization goals\cite{nrelgreen}.
For instance, in the  ``{100\% annual renewable matching}'' program\footnote{RE100 Initiative: \url{https://www.there100.org/}.}, users attempt to purchase the same amount of renewable energy as the electricity it consumes in a year. To enhance the matching granularity, the strategy of ``{24/7 matching}'' or ``{24/7 carbon-free energy}'' is proposed, aiming to match the purchaser's electricity demand on an hourly basis (24 hours/7 days) with renewable generation in the same power grid region. 
A growing number of  entities\cite{247comp},  such as Google, Microsoft, the U.S. federal government\footnote{Executive Order 14057 - Catalyzing Clean Energy Industries and Jobs Through Federal Sustainability,
December 8, 2021.}), are committed to procuring 24/7 carbon-free electricity \cite{xu2472021}.  
Reference \cite{xu2472021} analyzes the system-level impacts of 24/7 carbon-free electricity procurement, and recent studies \cite{miller2020beyond, mic247} propose implementation frameworks for  24/7 renewable energy procurement. See the white paper\cite{247epri2022} for a comprehensive  review of the 24/7 carbon-free energy programs and industry practice.

\vspace{2pt}
 
\noindent \textbf{Discussions.} 
A significant advantage of contractual instruments and market-based carbon accounting is that they enable corporations to proactively manage their carbon emission footprints, rather than passively taking grid average emission factors for carbon accounting, over which they have limited control. Therefore, market-based accounting mechanisms offer
opportunities and incentives to engage vast users in the process of decarbonization. Nevertheless, it is controversial to use market instruments for carbon accounting\cite{brander2018creative, bjorn2022renewable, bjorn2022rec}, and we summarize the critical issues below.
\begin{itemize}
 \item [1)] (\emph{Disconnect with Physical Grids}). Most existing market-based methods are purely contractual and do not consider physical power grid operation (violation of Principle \ref{prin:physical}). For example, a company can claim zero carbon emissions by simply procuring RECs, even though the renewable generation does not coincide with its electricity use. Recent studies  \cite{bjorn2022renewable,bjorn2022rec,monyei2018electrons} imply that corporate REC purchases may not substantially incentivize additional  renewable generations but allow companies to overstate their emission reductions. To this end, advanced market-based instruments and programs are under development to align carbon accounting with physical systems to ensure \emph{deliverability} of clean electricity \cite{ricks2023minimizing}.
 
    \item [2)] (\emph{Residual Mix}). Another critical issue is how to allocate
 the unclaimed emissions, known as \emph{residual mix} \cite{world2014scope2}, to the entities that are not involved in contractual instruments. In current practice,
only clean power purchasers report reduced carbon emissions, but no one else takes responsibility for increasing emissions in the residual mix (violation of Principle \ref{prin:comm}). Additionally, accurately estimating the residual mix emissions is challenging despite some early efforts\cite{resi2021,greene},
as it requires tracking all transactions of contractual instruments.



 \item [3)] (\emph{Implementation Challenges}).
   Market-based carbon accounting presents various practical implementation challenges, including  tracking all power purchase and sale contracts, defining proper market boundaries for accounting,
potential double accounting issue,  and lack of regulations, standards, and unified tracking and management platforms, etc. 
\end{itemize}

\subsubsection{Consequential Carbon Accounting }\label{sec:avoid}

In contrast to
\emph{attributional} carbon accounting above, 
  \emph{consequential} carbon accounting aims to 
   measure and quantify the emission consequences (the \emph{avoided} or \emph{induced} emissions) for a specific electric decision or project, such as performing demand response and deploying new renewable generators.
      Essentially, consequential accounting identifies the causal relationship between an electric activity and the resultant carbon emission change, compared to a {counterfactual baseline scenario} (``\emph{baseline emissions}") \cite{broekhoff2007guidelines} in which the activity does not occur. Therefore, consequential carbon accounting can also serve as a useful tool to inform decarbonization decision-making \cite{greg2022conse}.

To facilitate the evaluation of avoided/induced emissions, the \textbf{Marginal Emission Factor (MEF)}  is used to quantify the emissions impact associated with a marginal change in electricity supply or demand. There exist various definitions of MEF for different use cases and timescales, such as operating margin, build margin, long-term margin, and short-term margin \cite{siler2012marginal,greg2022conse}.
In addition, a variety of approaches have been proposed to estimate MEF based on different assumptions and datasets, including non-base load methods \cite{rothschild2009total,broekhoff2007guidelines}, simulation methods \cite{zheng2015assessment,wang2014locational}, merit-order methods \cite{deetjen2019reduced,baumgartner2019design}, price-based methods \cite{rogers2013evaluation}, statistical methods \cite{wang2016estimating,corradi2018estimating}, etc. There are multiple platforms providing marginal emission data, such as NREL Cambium, EPA eGRID, EPA AVERT, Singularity Energy, WattTime \cite{wattime2022}; see the guidance \cite{greg2022marginal} for a summary of available marginal emission data sources.

Nevertheless, it is debatable to use MEF for consequential carbon accounting due to three key issues: 
1) MEF is often calculated as the generation emission factors of the marginal generators in the power grid, but as noted in the PJM's Primer \citep{mefpjm}, 
  marginal units and the associated MEFs
    can not predict future emission consequences. 
    Because, in the electricity market, the calculation of the local marginal price (LMP) identifies the
    marginal units, which change every five minutes and thus can not provide consequential predictions. 2) Marginal emission methods generally make the assumption that only one entity is taking action while others remain unchanged, which does not hold in practice, making it intractable to accurately evaluate the emission impact of a particular activity. It may also give rise to a situation where multiple entities lay claim to the same emission reduction credits (violation of Principle \ref{prin:comm}). 3) 
    MEF focuses on marginal incremental changes and thus may not effectively encompass the impact of varying magnitudes of change. 
    For example, 
a 1-watt load increase and a 100MW load increase will trigger distinct generation responses, leading to different consequential emission factors.

\begin{figure}
    \centering
    \includegraphics[scale =0.65]{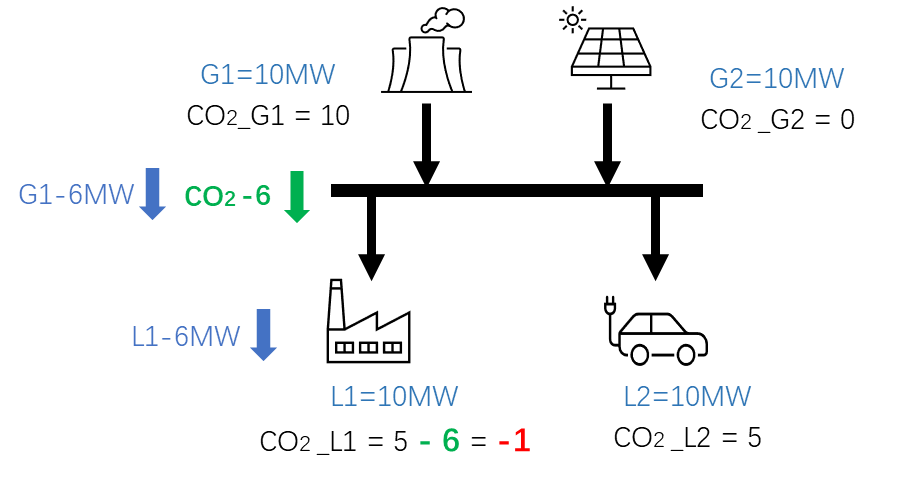}
    \caption{An illustrative example of avoiding the mixing of attributional and consequential carbon accounting. (Suppose that there are two 
     generators (G1 and G2) and two loads (L1 and L2) in the power system.
     G1 is coal-fueled and produces 10 MW of electricity power with 10 units of carbon emissions per hour, while G2 is solar and produces 10 MW of electricity with zero emissions. Both L1 and L2 consume 10 MW of electricity power, and thus the attributed carbon emissions to each of them are 5 units. Then, L1 performs demand response and reduces its load by 6 MW, which results in a 6MW generation reduction in G1 along with the avoided emissions of 6 units. 
    If L1 uses the avoided emissions
    for offset, the emissions for L1 will be $5-6 = -1$ unit. This result is invalid, as L1 still consumes 4MW of electricity power but reports negative emissions.) }
    \label{fig:avoid}
\end{figure}

\noindent
 \textbf{Discussions.} We note that consequential carbon accounting is generally \emph{incompatible} with attributional carbon accounting. It means that one can not use avoided emissions to offset attributional emission inventories, which can lead to inconsistency and double accounting. To illustrate, consider the simple example shown in Figure \ref{fig:avoid}, where the mixing use of these two types of accounting schemes leads to invalid carbon emission results.
 In addition, accurately predicting, assessing, and verifying 
consequential emission impacts are challenging due to the intricate nature of power systems and the need to compare them with \emph{counterfactual} baseline emissions.

Moreover, consequential carbon accounting should be temporally and spatially granular (Principle \ref{prin:resolution}) and take into account the physical power grid operation (Principle \ref{prin:physical}). To highlight the importance of factoring in physical power flow and network constraints, we present a counter-intuitive example later, as depicted in Figure \ref{fig:congest}, where an \emph{increase in electricity consumption} leads to a \emph{decrease in the grid's emissions} due to power line congestion. The concept of \emph{locational marginal emissions} \cite{lmeresurety,wang2014locational} is introduced to consider the geographical diversity and power grid constraints when evaluating the marginal emission impacts.

\subsection{Flow-Based Carbon Accounting Framework} \label{sec:cef}

As discussed above, most existing attributional and consequential carbon accounting methods do not take into account the physical power grids, underlying power flow, or network constraints (violation of Principle \ref{prin:physical}). Such methods are commonly referred to as ``\emph{pool-based}'', since they treat the power system as an electricity pool without consideration of power flow delivery. These pool-based carbon accounting methods often encounter a disconnect from physical grids and may fail to accurately reflect the physical relation between electricity consumption and emissions produced on the generation side. This can result in unfaithful and uninformative carbon accounting outcomes and may mislead decarbonization projects, investments, market design, and policy development. For instance, 1) it is inappropriate for a data center that is completely powered by a nearby wind farm to use the grid AEF for carbon accounting; 2) the counter-intuitive scenario depicted in Figure \ref{fig:congest} cannot be accounted for when disregarding power flow delivery and power grid operational constraints.

To address these issues, this paper proposes a comprehensive \emph{flow-based carbon accounting} framework that aligns carbon accounting with physical power grids and power flow. This flow-based framework is applicable to both attributional and consequential carbon accounting. In addition, pool-based accounting methods can be viewed as a reduced version of this framework. In the following subsections, we elaborate on the 
concept and methods of flow-based attributional and consequential carbon accounting, as well as critical application issues.

\subsubsection{Flow-Based Attributional Carbon Accounting} \label{sec:flowatt}

\begin{figure}
    \centering
        \includegraphics[scale=0.53]{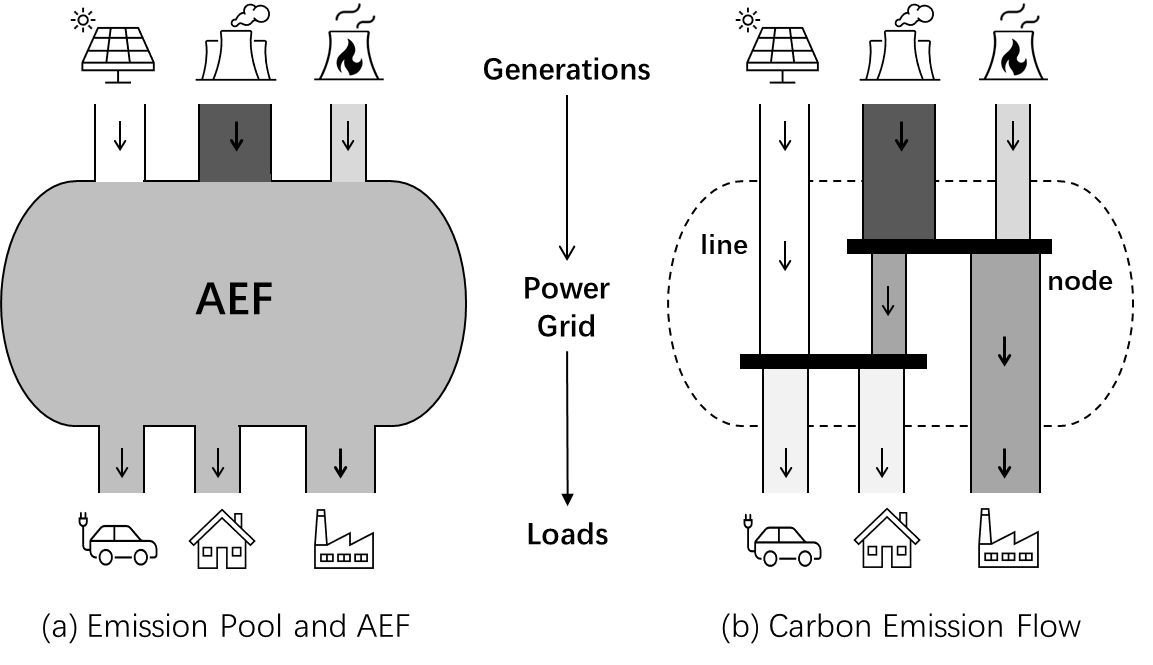}
    \caption{Comparison between carbon emission pool with AEF and carbon emission flow. (The width of each pipeline denotes the magnitude of power flow, and darker colors indicate higher carbon emission intensities.)}
    \label{fig:pool}
\end{figure}

  \textbf{Concept of Carbon (Emission) Flow}. To align demand-side carbon accounting with physical power flow, carbon emissions generated by power plants can be viewed as \emph{virtual} attachments to the power flow. These emissions are considered to be transmitted from  generators through  power grids and accumulate on the demand side, forming virtual \emph{carbon (emission) flow} over the power network. 
It is helpful to comprehend the concept of carbon flow using the analogy of ``water flow'', as illustrated in Figure \ref{fig:pool}.
Specifically, the power flow from a generator through a power line is analogous to the water flow from a water  source through a pipeline.
The virtual carbon emissions accompanying  power flows can be viewed as invisible particles contained in   water flows, and the  {carbon emission intensity} is analogous to the particle concentration. A renewable energy generator with zero-emissions is analogous to a pure water source that supplies zero-particle water. 
Similar to water flows with different particle concentrations blended together at a water station, power flows with different carbon intensities are assumed to mix at each node of the power network. The concept of carbon   flow was first introduced in \cite{atkinson2011trade,iniCEF}. References   \cite{kang2012carbon,kang2015carbon} propose the carbon emission flow in power networks and established its mathematical models. 
It is also worth noting that the carbon flow method is closely related to the long-standing issue of \emph{electricity tracing} (or power flow tracing) \cite{chen2019tracing,chang2001electricity} in power systems, which dates back to the 1990s \cite{bialek1996tracing} and 
was initially studied for electricity market operation. Electricity tracing identifies which generator supplies which load, making it useful for carbon emission footprints tracing in power grids.


\vspace{2pt}

\begin{figure}
    \centering
    \includegraphics[scale=0.4]{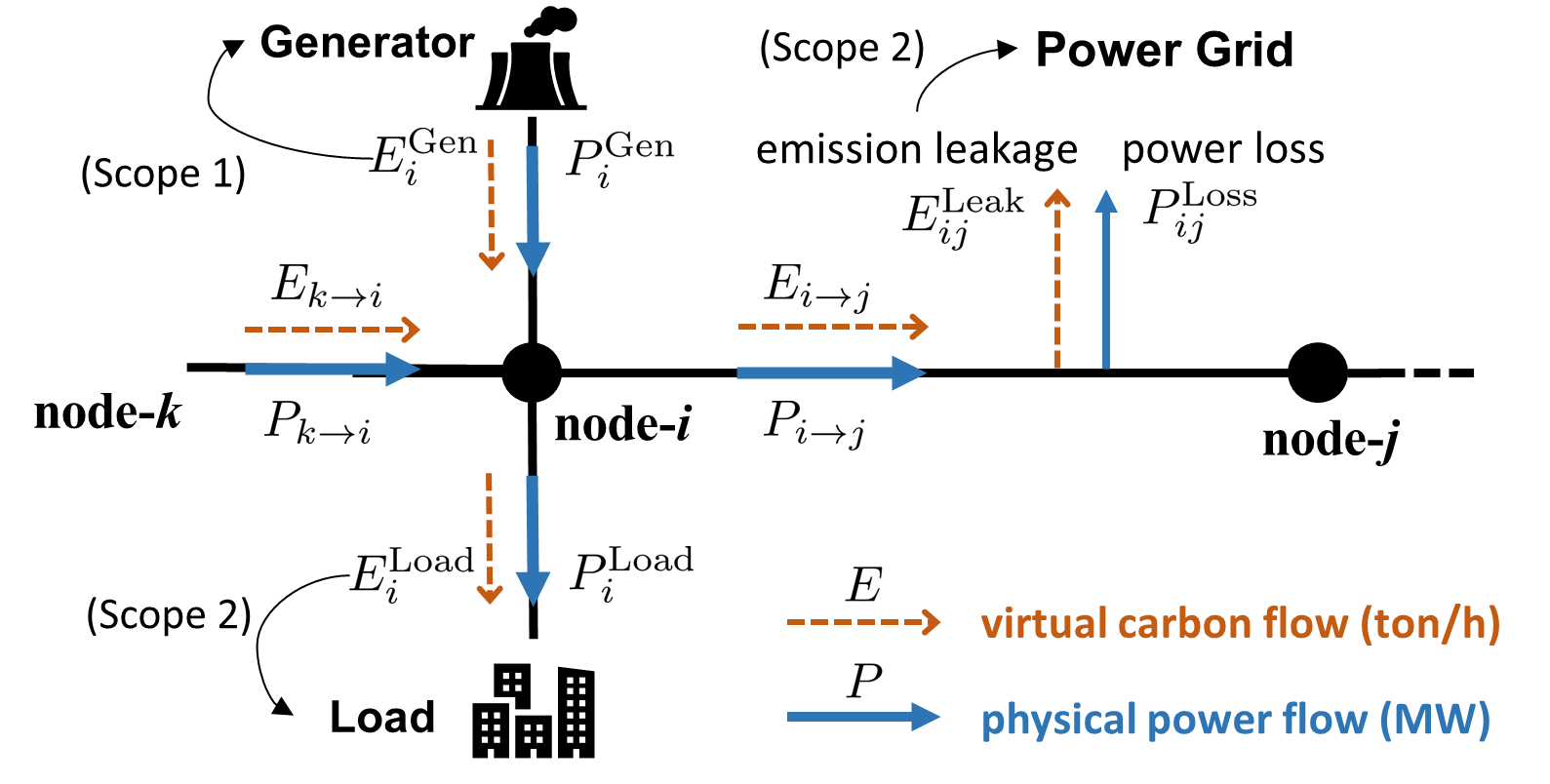}
    \caption{Physical power flow and virtual carbon  flow in a power network. ($P_i^{\mathrm{Gen}}$ 
    and $E_i^{\mathrm{Gen}}$ denote the generation power and the produced carbon emissions at node $i$.
    $P_i^{\mathrm{Load}}$ and $E_i^{\mathrm{Load}}$ denote the load consumption and the associated demand-side carbon emissions at node $i$. 
 $P_{i\to j}$ and $E_{i\to j}$ denote the physical power flow and virtual carbon flow of line $ij$ from node $i$ to node $j$. 
 $P_{ij}^{\mathrm{Loss}}$ and $E_{ij}^{\mathrm{Leak}}$ denote the power loss and associated carbon emission leakage of line $ij$.) 
}
    \label{fig:cfmodel}
\end{figure}

\noindent
  \textbf{Carbon Flow Model}.
Figure \ref{fig:cfmodel} illustrates the physical power flows and the virtual carbon flows in a power network. 
The notations with $P$ denote  power flows (in the unit of MW), while the notations with $E$ denote the associated carbon emission flows (in the unit of ton/h). In addition, the notion of \textbf{carbon emission intensity} $w$ is defined as $w = E/P$ to link power flows and carbon flows. This measures the amount of carbon emissions  associated with 1 unit of power, 
which is similar to but more granular than AEF.
The carbon flow model is then established based on the following two fundamental rules.

\begin{rules}   
    ({Conservation of Carbon Mass}). At each node, the total carbon emission inflows equals the total carbon emission outflows. 
\end{rules}   

\begin{rules}   
    The power flow and power loss of each power line have the same carbon emission intensity. 
\end{rules}

Moreover, a \emph{carbon flow mixing rule} is generally needed to route total carbon emission inflows to all outflows at each node. One widely-used mixing rule is ``\emph{proportional sharing}'' \cite{kang2015carbon, bialek1996tracing}, which assumes that the total carbon emission inflow is distributed among outflows in proportion to their power flow values. 
Equivalently, the proportional sharing rule indicates that all outflows at each node share the same \emph{nodal carbon intensity}. Under these rules and given power flow values, carbon flows and nodal carbon intensities across the power grid can be computed. See references \cite{chen2023carbon, kang2015carbon} for more details on the mathematical model of carbon flow.




 \vspace{2pt}
\noindent
 \textbf{Use of Carbon Flow for Attributional Carbon Accounting}. Note that carbon flows vary over time due to changes in generation fuel mix and time-varying power flow. We take the case 
 in Figure \ref{fig:cfmodel} for example and consider a time period $\mathcal{T}=\{1,2,\cdots,T\}$ with time resolution $\Delta t$. In compliance with the GHG Protocol \cite{world2014scope2}, the flow-based attributional carbon accounting framework implies that 
 1) the generator at node $i$ shall report the Scope 1 carbon emissions of $\sum_{t=1}^T\Delta t\cdot E_{i}^{\mathrm{Gen}}(t)$, 2) the load at node $i$ shall report the Scope 2 emissions of $\sum_{t=1}^T\Delta t\cdot E_{i}^{\mathrm{Load}}(t)$, and 3) the power grid owner shall report the Scope 2 emissions of $\sum_{t=1}^T\big(\Delta t\cdot\sum_{ij}  E^{\mathrm{Leak}}_{ij}(t)\big)$ associated with power loss. In this way, the carbon accounting results are built upon physical power flows and can achieve high spatial granularity (each node has its own nodal carbon intensity) and high temporal granularity (the time resolution $\Delta t$ can be 1 hour or even 1 minute).




\begin{remark}
\normalfont We note that
the proposed flow-based attributional accounting framework above is generic and flexible:
\begin{itemize}
    \item [1)] The \emph{definition of ``node'' is flexible}, which can be the connection point of an electric facility, a district grid, a distribution grid, or a bulk transmission system, depending on practical requirements and data availability. In particular, when taking the entire balancing area as a single node, the flow-based accounting method reduces to the pool-based carbon accounting method with AEF. 
    
    \item [2)] The \emph{carbon flow mixing rule at each node can be flexible}. In addition to the commonly used proportional sharing rule, alternative sharing or mixing rules can also be applied depending on practical use cases. In particular, when considering no power inflow mix at all, the flow-based accounting framework has the potential to \emph{align contractual energy paths with the physical power flow} and offer physical implications for market-based instruments such as power purchase agreements. 
See  Section \ref{sec:comparison} and  Figure \ref{fig:poolflow} for more discussions. 
\end{itemize}

\end{remark}



\subsubsection{Flow-Based Consequential Carbon Accounting} \label{sec:f-consequential}

\begin{figure}
    \centering
    \includegraphics[scale=0.36]{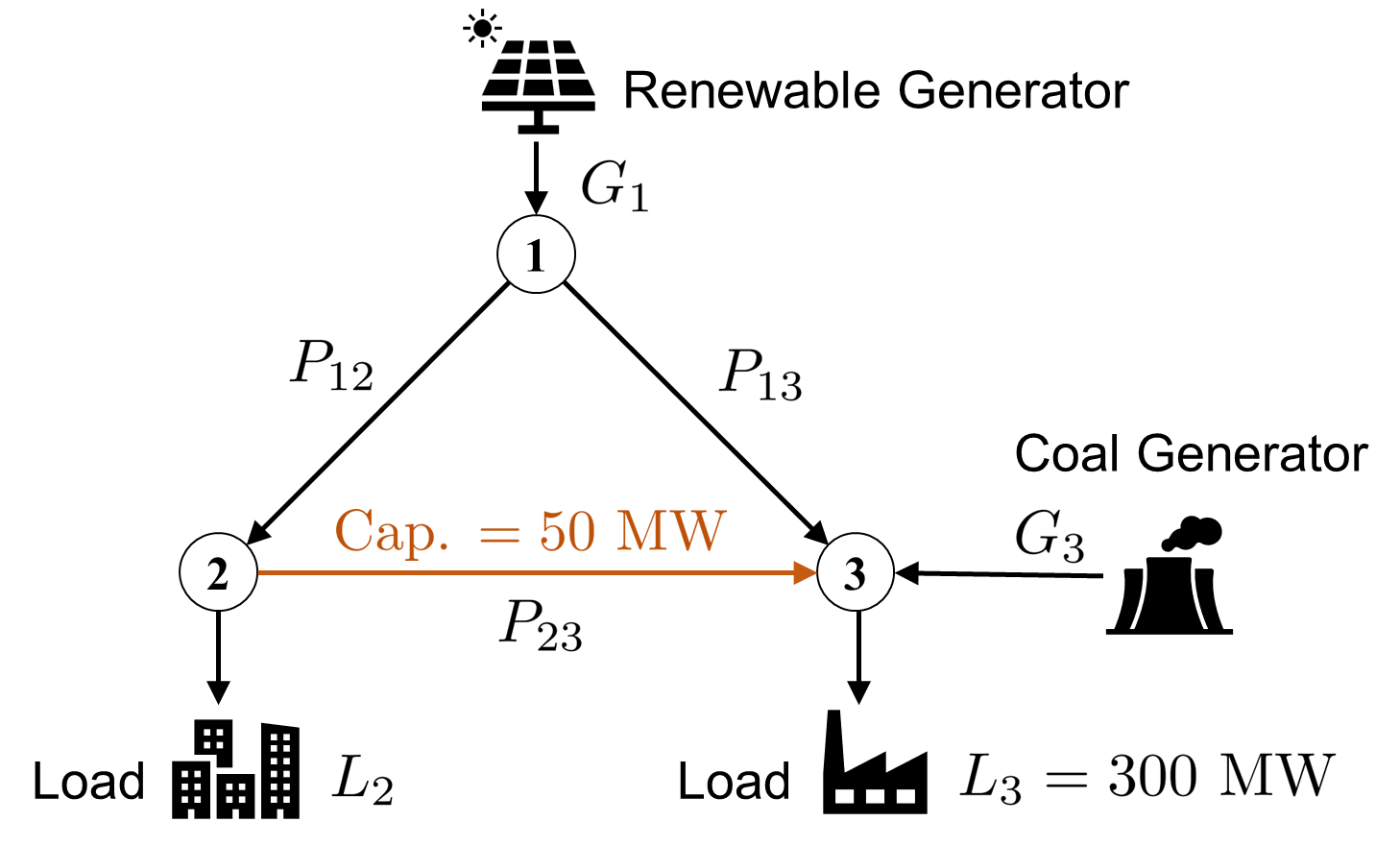}
    \caption{A three-node power grid, with a renewable generator $G_1$ at node-1 and a coal generator $G_3$ at node-3 whose generation emission factor is 1 unit \coo/MW. There are two loads $L_2, L_3$ at node-2 and node-3 with $L_3=300$MW. Every power line has the same line impedance. Let 
    $P_{ij}$ denote the power flow of line $ij$ from node $i$ to node $j$. Line 23 has a power capacity of 50 MW and may be congested. Assume that other power lines and generators have sufficiently large power capacity limits. The power grid's carbon emissions can be reduced from 150 units to zero by increasing load $L_2$ from 0 to $150$ MW, due to power line congestion.
    } 
    \label{fig:congest}
\end{figure}

As indicated by Principle \ref{prin:physical}, the physical power grid operation and network constraints, such as line capacity limits and safe voltage constraints, 
are pivotal factors in determining the consequential emission impacts of an electric activity or project. 
To illustrate it, we present a counter-intuitive example, as shown in Figure \ref{fig:congest}, where \emph{an increase in electricity consumption paradoxically leads to a reduction in total system emissions} due to power line congestion. 

Specifically, for the three-node power grid in Figure \ref{fig:congest}, consider the case when the load $L_2$ increases from 0 to $150$ MW. When $L_2 =0$ MW, according to the physics laws (namely Kirchhoff’s laws and Ohm’s law), the power flows must satisfy $P_{13} = 2 P_{12} = 2P_{23}$, as each power line has the same impedance. It implies that in order to transmit 1MW of power from node 1 to node 3 through line 13, there must be an accompanying 0.5 MW of power transmitted through line 12 and line 23. However, since line 23 has a power capacity of 50 MW, at most 150 MW of carbon-free electricity can be transmitted from the renewable generator $G_1$ to load $L_3$, namely $P_{13}=100$ MW and $P_{12} = P_{23}=50$ MW. Then, the rest 150 MW of load $L_3$ needs to be supplied by the coal generator, resulting in 150 units of carbon emissions. When increasing load $L_2$ to $150$ MW, the power flows satisfy $P_{13} = 2P_{23} + L_2$ according to physics laws. Then, at most 300 MW of carbon-free electricity can be transmitted from $G_1$ to $L_3$ with $P_{13} = 250$ MW, $P_{12}=200$ MW, and $P_{23}=50$ MW. As a result, no electricity supply is needed from the coal generator, and \emph{the system's carbon emissions are reduced from 150 units to zero by increasing load $L_2$ from 0 to $150$ MW}.

The illustrative example above demonstrates the significance of power grid congestion in determining the consequential carbon emission impacts. 
To assess the impact on congestion 
with respect to the changes in nodal power injection (load or generation), the
Power Transfer Distribution Factors (PTDFs) \cite{chao2000flow}, a well-known concept in power system engineering, can serve as a useful tool. 
PTDFs describe the incremental change in line power flows caused by a power injection at a (source) node and the same amount of power withdrawal at another (sink) node. In other words, PTDFs provide a linearized approximation of how the power flows on transmission lines change in response to a transaction between the electricity seller (source) and the buyer (sink). The concept of Carbon-PTDF (C-PTDF) that integrates carbon emission impact to PTDF can be useful for consequential carbon accounting with respect to the network constraints.

Essentially, to assess the consequential emission impact of an electric action ${x}$, one needs to simulate the power system frequency response or solve the optimal power flow (OPF)-based economic dispatch to obtain the change in the generation profile $\Delta \bm{G}$. This response process can be described as a mapping function $\Delta\bm{G} = \bm{g}({x})$. The consequential emission change $\Delta E$ is then calculated based on the generation profile change as $\Delta E = \bm{w}_G^\top \Delta\bm{G} = {f}({x})$, where $\bm{w}_G$ denotes the associated generation emission factors. The function $f$ depicts how the electric action $x$ affects the system emissions. However, this function $f$ is highly complex as it captures the entire power grid information and is generally unknown to end users. The MEFs  introduced in Section \ref{sec:avoid} are essentially the simplifications of the function change $\frac{\Delta f}{\Delta x}$ for practical use.

At last, it is worth noting that (flow-based) attributional carbon accounting and consequential carbon accounting are closely related, despite being designed for distinct purposes. Basically, 
attributional accounting describes the grid emission status in a steady state, while
consequential accounting depicts the change of grid emission status over time. The mathematical relation between attributional emission results and consequential emission results can be established based on the generation response and power flow models.

\subsubsection{Comparison between Flow-Based and Pool-Based Methods} \label{sec:comparison}


\begin{figure}
    \centering
    \includegraphics[scale=0.4]{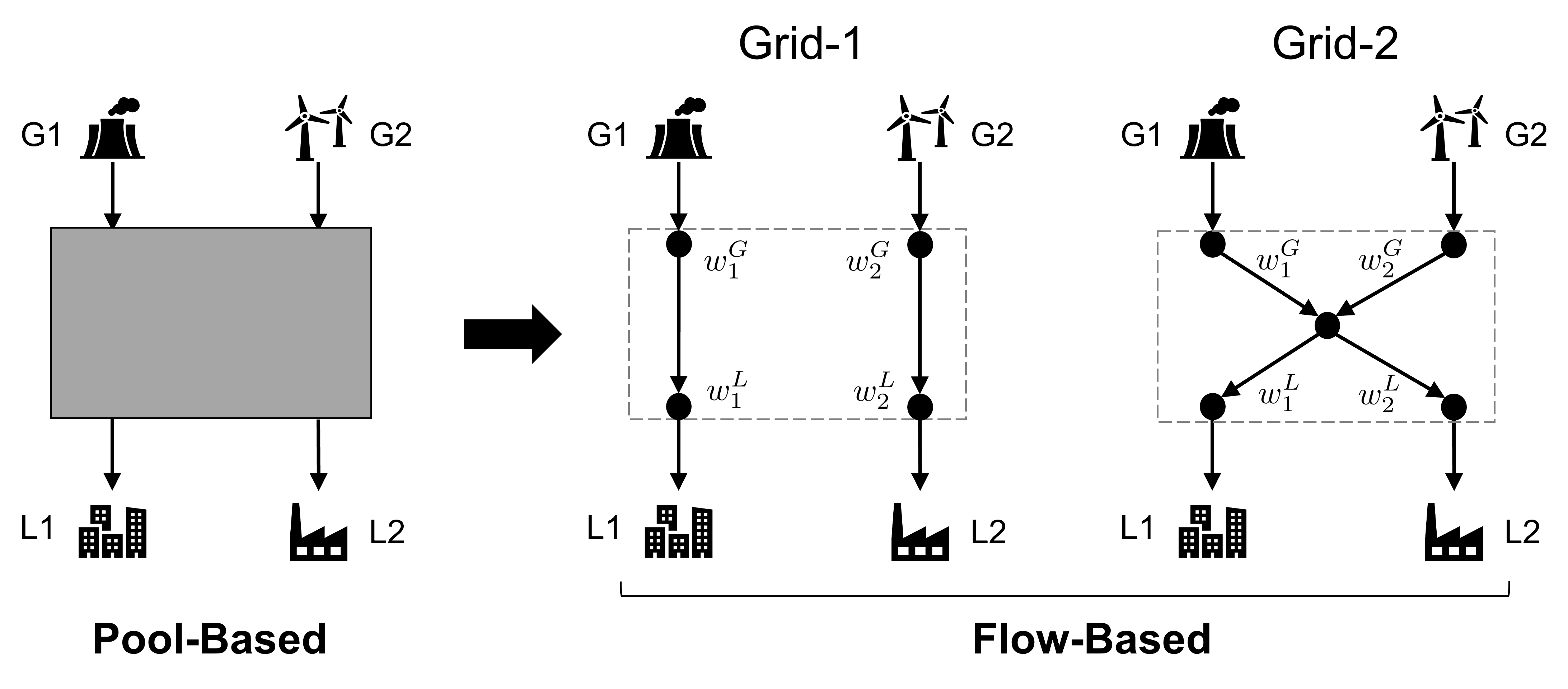}
    \caption{Comparison between pool-based and flow-based carbon accounting. (There are a fossil-fueled generator (G1), a renewable generator (G2), and two loads (L1 and L2) in the system. In contrast to pool-based accounting, flow-based accounting methods consider specific power grid topology and power flow, and different grid topologies lead to different accounting results. In Grid-1, under flow-based accounting, the emission intensities for Load-1 and Load-2 are calculated as $w_1^L \!=\! w_1^G$ and $w_2^L \!=\! w_2^G$, respectively. In Grid-2, the conservation law of mass for carbon flow gives that $w_1^G G_1 + w_2^G G_2 \!=\! w_1^L L_1 + w_2^L L_2$; when the proportional sharing rule is adopted, we further have  $w_1^L = w_2^L = ({w_1^G G_1 +w_2^G G_2})/({L_1 + L_2})$. Then, in both two cases, the attributed carbon emissions for Load-1 and Load-2 are calculated as $w_1^L L_1$ and $w_2^L L_2$, respectively. }
    \label{fig:poolflow}
\end{figure}

In contrast to pool-based accounting approaches, flow-based carbon accounting methods are well aligned with the physical power grid and its power flows. Moreover, flow-based consequential accounting methods also consider the operational constraints of the power system, such as line congestion limits, providing more granular and informative accounting results. More recently, flow-based accounting has gained traction in the industry, encompassing both attributional \cite{totalCA2021, de2019tracking} and consequential \cite{mefpjm, lmeresurety}) accounting.

We further explain the advantages of flow-based accounting below, using the simple case in Figure \ref{fig:poolflow} as an illustrative example.

\begin{itemize}
    \item[1)] By aligning with physical power flows, flow-based accounting can resolve many unjustifiable issues that pool-based accounting encounters. For example, 
     in Grid-1 shown in Figure \ref{fig:poolflow}, Load-2 (L2) is entirely supplied by renewable Generator-2 (G2). In flow-based accounting, the carbon emission intensity for Load-2 is $w_2^L = w_2^G = 0$, and Load-2 reports zero emissions. This is consistent with the physical electricity supply and fairer
     than using the grid average emission factor in pool-based accounting. 
  \item[2)] Flow-based accounting can rule out physically unrealistic power purchase agreements or contractual claims, which pool-based accounting may fail to identify.
   For instance, as shown in Figure \ref{fig:poolflow},
  carbon-free electricity can not be delivered from Generator-2 to Load-1 in Grid-1 due to no power line connection, but it is feasible in Grid-2. 
Moreover, flow-based accounting methods can be used to assess the \emph{physical deliverability} of power purchase agreements or contracts, which is a critical issue of surging attention for many large energy consumers that engage in contractual instruments.
    
 \item[4)] 
 By designing the carbon flow mixing rule, flow-based accounting has the potential to provide physical implications for market-based instruments. For example,
 in Grid-2 (in Figure \ref{fig:poolflow}), if there is an electricity purchase contract from renewable Generator-2 to Load-1, instead of using the proportional sharing rule, one can define a new mixing rule that forces the carbon emission intensity $w_1^L = 0$ for Load-1, and then the carbon intensity for Load-2 is calculated as 
  $w_2^L \!=\! (w_1^G G_1 + w_2^G G_2)/L_2$ under the conservation law of mass, which is larger than the carbon intensity $(w_1^G G_1 + w_2^G G_2)/(G_1+G_2)$ computed using the proportional sharing rule, leading to more attributed emissions for Load-2. Therefore,
  this approach potentially provides a systematic way to calculate and allocate the residual mix.


    \item [5)]  Flow-based carbon accounting can produce granular emission results, which enables clear visualization of the carbon footprints and geographical travel directions in wide areas. This provides useful information and insights for system-level emission analysis and decarbonization  decision-making.
    

\end{itemize}


\subsection{Key Issues} \label{sec:discussion}


This subsection highlights the key issues that generally arise in carbon accounting.
One of the fundamental problems is that 
electrons are indistinguishable physically, but the carbon accounting attempts to differentiate electricity by assigning it a carbon label. An important question behind carbon accounting is that 
``\emph{whether electricity is fungible or traceable}".
The answer to this question can significantly influence the design philosophy of carbon accounting mechanisms. 
Several other critical issues are discussed below:
\begin{itemize}
    \item (\emph{Adaptive Granularity}). 
Enhancing the spatial and temporal granularity of carbon accounting generally provides more detailed and informative emission results, albeit with the trade-off of increased demands on data and effort. An adaptive approach can be employed in selecting proper granularity for specific use cases, time scales, and requirements.

    \item (\emph{Data Availability and Disclosure}). End-users generally lack access to detailed power system operation information. To facilitate effective and physics-aligned carbon accounting, it is essential for power plants, Independent System Operators (ISOs), and utilities to disclose accurate and granular grid emissions data in a timely manner, for example, by providing real-time nodal carbon emission intensity data to users.

    \item (\emph{Equity and Fairness}). In the design of carbon accounting mechanisms, equity and fairness are 
    critical factors. Different methods can result in significantly varied emission allocations among end-users, with these allocations reflecting real-world social responsibility and economic consequences. As such, carbon accounting mechanisms should take into account fair distribution of emissions across various groups with distinct socio-demographic characteristics.

\end{itemize}

%% file: Decision.tex
\section{Carbon-Aware Decision-Making}
\label{sec:decision}

To combat climate change, electric power systems are undergoing an architectural transformation to attain sustainability and carbon neutrality under the strategy of ``high renewable penetration + deep electrification''. On the one hand, fossil fuel-based (such as coal, natural gas, oil) power plants are being replaced by renewable energy sources (such as solar and wind generation).
On the other hand, 
 deep electrification 
 is taking place on the demand side, 
 especially within the transportation and building sectors \cite{SCEpathway}, including the replacement of gasoline-powered vehicles by electric vehicles and the replacement of natural gas heaters by electric heating systems. However, these substantial transformations introduce a series of emerging challenges for power systems, spanning the \emph{generation}, \emph{grid}, and \emph{demand} layers: 
 \begin{itemize}
    \item  (\textbf{Generation Layer}): 
Wind and solar generations are different from conventional thermal generators as their power outputs depend on local weather conditions. As a result,
    high penetration of renewable energy sources brings significant variability and uncertainty to power systems. 
    
        \item  (\textbf{Grid Layer}): Modernized and resilient power grids with sufficient transmission capacity are necessary to accommodate
          growing loads and intermittent renewable generation, while existing grid infrastructures are generally not ready to support carbon-neutral electricity systems
          \cite{xie2021toward}.
          
            \item  (\textbf{Demand Layer}): 
            Deep electrification will lead to huge increases in electricity demand and sharpen peak loads. 
            Coordinating massive load facilities and guiding users in their electricity consumption is essential to ensure reliable system operation and expedite the process of decarbonization.

            
            
    
\end{itemize}

Built upon the foundation of carbon accounting in Section \ref{sec:accounting}, this section aims to explicitly integrate the context of carbon emissions into the planning, operation, and control decision-making of electric power systems. Firstly, we introduce the fundamental methodology of Carbon-aware Optimal Power Flow (C-OPF) for grid decarbonization decision-making. Then, we frame 
the major decision-making problems across the layers of generation, grid, and demand  at different time scales. Note that the division of these three layers is  mainly to facilitate the discussion, while the   decision problems in power systems generally interconnect all three layers to achieve global optimality. The carbon emission models of energy storage systems and several critical issues on engaging end-users' actions for decarbonization are also discussed.

\subsection{Carbon-Aware Optimal Power Flow (C-OPF)} \label{sec:copf}

The optimal power flow (OPF) method \cite{frank2012optimal,cain2012history} is a foundational and  widely-used tool  for power system planning, 
operation, control, and market. 
OPF develops an optimization model that seeks to optimize decisions (for example, power generation or network topology) to minimize a given cost objective while satisfying network power flow constraints,  equipment operating limits,  reliability requirements, etc. 
Power system decisions directly affect carbon emissions, and different power flows generally result in different carbon footprints and attribution outcomes for users. 
However, the existing OPF schemes do not explicitly take  into account carbon emissions. To this end, 
built upon the classic OPF method, we propose the Carbon-aware Optimal Power Flow (C-OPF) model, formulated as \eqref{eq:copf}, which incorporates the carbon flow equations and  constraints, as well as carbon-related objectives to enable the co-optimization of both electric energy flow and carbon emission flow. 
\begin{subequations} \label{eq:copf}
\begin{align}
\text{Obj.}  \ \ &  \min_{\bm{x} \in \mathcal{X}} \  C_{\mathrm{power}}(\bm{x},\bm{y}) + C_{\mathrm{carbon}} (\bx,\by,\bm{z}) \label{eq:copf:obj}\\
 \text{s.t.} \ \   & \text{Power Flow Equations }(\bm{x}, \bm{y}) = \bm{0}\label{eq:copf:pf}\\
    & \text{Power Flow Constraints } (\bm{y})\leq \bm{0} \label{eq:copf:pfc}\\
        & \text{Carbon Flow Equations } (\bm{x}, \bm{y}, \bm{z}) = \bm{0}\label{eq:copf:cf} \\
    & \text{Carbon Flow Constraints } (\bx, \by, \bm{z})\leq \bm{0}. \label{eq:copf:cfc}
\end{align}
\end{subequations}
Here, $\bx$ denotes the grid decision variable that is subject to the feasible set $\mathcal{X}$. 
$\bm{y}$ denotes the power flow variable and $\bm{z}$ denotes the carbon flow variable. For example, in the economic dispatch problem, $\bx$ can be the generation decisions of generators and $\sX$ describes the generation capacity limits. $\by$ captures the voltage profile and power flows over the network.   $\bz$ represents the carbon emission flows and nodal/branch carbon intensities. The objective \eqref{eq:copf:obj} aims to minimize the total cost, including 
  the cost associated with power flow (e.g., generation cost) and the cost associated with carbon emissions (e.g., regulation penalty on carbon emissions). \eqref{eq:copf:pf} denotes the power flow equations and \eqref{eq:copf:pfc} describes the power flow constraints, such as the voltage limits and line capacity constraints. \eqref{eq:copf:cf} denotes the carbon emissions flow equations that are introduced in Section \ref{sec:cef}, and \eqref{eq:copf:cfc} describes the constraints on carbon flow, such as caps on nodal carbon emissions \cite{cheng2018bi,wei2021carbon}, caps on  nodal carbon intensity  \cite{wu2022carbon}, 
  caps on the total cumulative emissions of a region,
  or equity requirement on emissions allocation \cite{sun2017analysis}.

 
 C-OPF can be viewed as a carbon-aware generalization of OPF. In particular, 
compared with conventional OPF schemes, our proposed C-OPF method has three key merits: 1) (\emph{Modeling Carbon Flow}): C-OPF explicitly models carbon flow alongside power flow, enabling a more accurate representation of the carbon emission footprints across the power system instead of only looking at system-wide total emissions. 2)  (\emph{Carbon Emission Constraints}): C-OPF imposes carbon emission constraints, ensuring that the system decisions adhere to emission reduction targets or regulatory requirements. and 3) (\emph{Carbon Accounting and Emission Optimization for End-Users}): C-OPF incorporates demand-side carbon accounting mechanisms and facilitates carbon-aware power dispatch decisions and allow users to actively manage their emissions.  
  Basically, the C-OPF method is applicable to existing OPF-based power system problems and solutions to enable carbon-aware decisions, including long-term expansion planning, short-term power scheduling (such as unit commitment,  economic dispatch, and demand response), electricity market design and pricing, and real-time control (such as frequency control and voltage control). See \cite{chen2023carbon} for more explanations on C-OPF. 

  



\subsection{Generation Layer: Power Plants}

Power plants in the generation layer are the primary sources of electricity and  direct carbon emissions in the grid.
We categorize the decarbonization decisions in the generation layer from three different timescales and discuss them below.

\subsubsection{Long-Term Planning}

One of the crucial long-term planning decisions for decarbonization in the generation layer is the optimal \textit{siting} and \textit{sizing} of \emph{new renewable generators} \cite{viral2012optimal}, especially wind and solar units. As renewable generations heavily rely on weather conditions, their uncertainty and variability need to be taken into account, in addition to the investment and operational cost, power network constraints, future load levels, and other factors. Depending on the types of renewable generations and stakeholders,  such as private  roof-top PV panels and large solar farms, the planning problems and schemes may differ significantly. In addition, 
the {retirement and replacement of fossil-fueled power plants} \cite{shen2020low}, such as
coal- and natural gas-fired power plants, is also critical for the transition to low-carbon electricity grids. 
In \cite{tao2020carbon}, a carbon emission flow-based method is employed to determine the construction site and capacity for  power-to-gas stations. The work
 \cite{cheng2018bi} on expansion planning of multiple energy systems considers the carbon emission constraints that the total carbon emissions of one energy hub during the planning periods should not exceed an emission cap. 

\subsubsection{Short-Term Operation} \label{sec:shortgen}

In terms of short-term operation, power plants participate in the electricity market and their generations are scheduled by Independent System Operators (ISOs) and utilities to meet the anticipated electricity demand through system-level least-cost power dispatch. 
The electricity generation from a power plant that runs on fossil fuels or nuclear energy is typically subject to the power capacity limits and ramping constraints, leading to time-coupled generation decisions. Conventionally, the coordination of various generators is achieved through the processes known as unit commitment (UC) \cite{padhy2004unit} and economic dispatch (ED) \cite{xia2010optimal}. UC involves scheduling which generators should start up or shut down and at what output levels, typically at a longer time horizon such as day-ahead planning. On the contrary, ED occurs at shorter intervals, such as every 15 minutes in a receding horizon fashion, and determines the output levels of the committed units to meet the real-time load demand. However, large-scale intermittent and uncertain renewable generations (such as solar and wind) introduce significant challenges to the grid 
in maintaining power balance between generation and demand. It requires advanced forecasting techniques, robust adaptive power dispatch schemes, and energy storage systems to facilitate the integration of renewable generators. 
Moreover, the negative externality of carbon emissions needs to be factored into the operation and dispatch of various generators, which can be internalized as a part of generation cost. In addition, 
 improving generation efficiency  and deploying carbon capture, utilization, and storage (CCUS) technologies \cite{zhang2020advances} are useful to decarbonize fossil-fueled power plants.



\subsubsection{Real-Time Control} \label{sec:gencontrol}

In real-time, generators are controlled over the timescale of seconds to minutes, to ensure the power balance between generation and demand and maintain the nominal grid frequency (e.g., 60 Hz in the U.S.)  through \emph{frequency control}.
Conventionally, power systems employ 
a hierarchical control structure consisting of three frequency control mechanisms operating at different timescales, including \emph{primary}, \emph{secondary}, and \emph{tertiary} frequency regulation, to achieve both fast response and economic efficiency. See \cite{bevrani2017intelligent} for detailed explanations of
the three-level frequency control architecture. Renewable generation units, such as wind turbines and solar panels, are  
inverter-interfaced with fast response capabilities but  lack physical or thermal inertia. As a result, large-scale integration of renewable generations  leads to lower system inertia, fast system dynamics, and greater control difficulty. Two types of control strategies for inverter-interfaced renewable units: \emph{grid-following} and \emph{grid-forming}  \cite{rosso2021grid,pattabiraman2018comparison}, are under active development.

\subsection{Grid Layer: ISOs and Utilities}

Electric power grids physically interconnect various power generators and numerous load facilities across wide geographic regions. And
power grid operation, generally managed by ISOs and utilities, plays a central role 
in delivering electricity from power plants to end-users.  
The C-OPF method introduced in Section \ref{sec:copf} is an useful tool to optimize the grid decisions to achieve operational efficiency and the goal of decarbonization.


\subsubsection{Long-Term Expansion Planning}

Due to the long lifespan of generation units, transmission lines, and other power facilities, it is difficult for the carbon emission level of a power grid to change significantly in a short period of time,
 which is known as the ``carbon lock-in" effect  \cite{chen2010power}. Hence,  optimal \emph{carbon-aware grid expansion planning} is critical for carbon emission reductions at scale  in the electricity sector. 
 This includes replacing aging power lines and electric equipment with higher capacity, less power loss alternatives, adding new transmission or distribution power lines to the electric network, and deploying new renewable generators and energy storage systems, among other measures.
In particular, the power network topology and configuration largely affect power flow and power loss over the grid and thus affect carbon accounting results. Therefore, power grid upgrade and expansion planning schemes can also incorporate carbon emission accounting and constraints based on the C-OPF method. 
In \cite{sun2017analysis}, the proposed transmission expansion planning scheme considers user-side carbon  accounting and defines a metric similar to the Gini index to promote ``carbon fairness'' across the grid.  
Moreover, energy storage facilities are generally required to accommodate large-scale uncertain renewable generations, and thus the \emph{siting and sizing of new energy storage systems} is  another key planning problem.

\subsubsection{Short-Term Operation} \label{sec:stps}

 As aforementioned in Section \ref{sec:shortgen}, in the short-term timescale (ranging from 
week-ahead, day-ahead, to 5 minutes ahead),  generation units are optimally scheduled to supply the demand in a reliable and efficient way through \emph{unit commitment} and \emph{economic dispatch},  
when 
the network constraints such as line capacity limits and voltage constraints 
are considered. 
To effectively reduce the system-level carbon emissions, the externality of carbon emissions and user-side emission footprints can be incorporated into the generation scheduling schemes by substituting the OPF method with the corresponding C-OPF method. In \cite{shao2019low}, a low-carbon economic dispatch   model is developed,
in which the penalty cost of demand-side carbon emissions is added to the objective function.
Reference   \cite{wang2021optimal}  proposes a low-carbon optimal scheduling model with demand response (DR) based carbon intensity control, which restrains the carbon intensities for all users to stay below a benchmark level.


Energy storage (ES)  systems are useful for power system decarbonization, as they can shift system-wise carbon emissions across time by switching between charging (power consumption) and discharging (power supply), potentially leading to time-coupled carbon accounting results. 
ES systems are expected to operate strategically by charging when the grid is primarily supplied by renewable generation and discharging when the grid relies on high-emission generation sources. 
The subject of carbon accounting for ES systems has attracted a great deal of recent attention from the industry \cite{esaccount2021,esacc2022}. 
In \cite{chen2023carbon}, two carbon emission models, together with the corresponding carbon accounting schemes, for ES systems have been proposed, including the ``\emph{water tank}'' model and the ``\emph{load/carbon-free generator}'' model. The former precisely models the dynamics of  virtually stored carbon emissions 
    and internal carbon intensity of ES units,  
    as well as carbon leakage associated with ES energy loss. The latter presents a simple model that treats ES as a load during charging and a carbon-free generator during discharging. 
See \cite{chen2023carbon} for detailed mathematical models.



The above discussions focus on the scheduling of active power, while the optimization of \emph{reactive power} are also critical in order
 to maintain safe voltage levels and reduce power loss in the grid. Therefore, carbon-aware reactive power optimization is needed to optimally schedule reactive power compensation devices for decarbonization. This is particularly relevant for power grid owners, as the carbon emissions associated with power loss constitute a significant share of their emission inventories.

\subsubsection{Real-Time Control}


In power grids, frequency level and voltage profile
are two of the most critical indicators of system operating
conditions. Accordingly, 
  \emph{frequency control} (explained in Section \ref{sec:gencontrol}) and \emph{voltage control} are two essential tasks to ensure the real-time stability and reliability of power systems.   
 In terms of voltage control \cite{chen2021model,sun2019review},  discretely controllable devices, such as on-load tap changing transformers,
voltage regulators, and capacitor banks, are generally scheduled in advance, as they are not suitable to be switched frequently. Continuously controllable devices, such as  inverter-based distributed energy resources (DERs) and
static Var compensators (SVCs), are then controlled in real-time to adjust their (active/reactive) power
outputs to maintain safe grid voltage levels under system disturbances.   
While
the current practice of power system control generally does not incorporate carbon emissions considerations, it is important to develop carbon-aware control schemes to 
fully exploit the potential for grid decarbonization.


\subsection{Demand Layer: End-Users}

Electrification of energy use, especially the transportation and building sectors, are taking place on the demand side to accelerate decarbonization. However, it will lead to huge increases in total electricity demand and greatly sharpen  peak loads, jeopardizing power balance and reliable system operation. To address these challenges, in addition to generation-side control, the coordinated management of massive load facilities and DERs on the demand side, such as data centers, smart buildings, smart homes, electric vehicles, and batteries, is necessary to exploit enormous user-side power flexibility  to accommodate deep electrification.





\subsubsection{Long-Term Planning}

 In the demand layer, long-term planning decisions include sitting new load centers, procuring clean energy, deploying DER facilities and infrastructure (such as electric vehicle charging stations), improving energy efficiency, and more.

\noindent \textbf{1) Sitting New Load Centers}.
Deciding the optimal location for new large load centers, such as energy-intensive factories or data centers, is a key issue in reducing carbon emissions for both the grid and the load.  Generally, 
 sitting a load center in proximity to renewable generators enables on-site utilization of clean energy and ensures a direct supply of clean power to the load. In \cite{6298199},  
 an optimization-based framework is proposed to minimize load center costs while considering carbon footprint reduction goals, renewable energy characteristics, and energy storage devices. The process of siting new load centers is a multifaceted problem that entails various considerations such as cost-profit analysis, power grid configuration, social impacts, and carbon emissions.


\noindent \textbf{2)   Clean Energy Purchase}. As discussed in Section \ref{sec:marketbase}, many large consumers purchase clean electricity from renewable energy suppliers, enabling them to report carbon emission reduction. Specific measures encompass purchasing renewable energy certificates (RECs), power purchase agreements (PPAs), implementing 24/7 carbon-free energy matching, etc. 
There are a number of studies \cite{pepper2022our,miller2020beyond} proposing the strategies to achieve the goal of 24/7 renewable energy. Nevertheless, it is necessary to evaluate the \emph{physical deliverability} of clean energy purchases and  incorporate it into the planning schemes.

\subsubsection{Short-Term Management} \label{sec:demand}

In short term, users can optimally manage
their loads in response to the needs of power system operation through load reduction, shifting, and shaping, namely  \emph{demand response (DR)} \cite{siano2014demand,chen2021online}.
In addition, the DERs owned by users, such as batteries, electric vehicles, rooftop PVs, and wind turbines, can be dispatched to adjust electricity consumption for decarbonization.
For instance, batteries can be scheduled to discharge during periods of high carbon emissions on the grid and charge when electricity is mainly supplied by renewable generation. Although a single DER device generally has a small capacity, the coordination of massive DERs can release huge power flexibility for grid reliability and decarbonization. Hence, this paper proposes 
the concept of 
\emph{carbon-aware virtual power plant} (C-VPP), where an aggregator coordinates a cluster of loads and DERs to operate collectively like a virtual power plant. 
This C-VPP can function as an entity for carbon accounting and reporting, actively engaging in system-level dispatch and market operations, and offering power flexibility to facilitate grid decarbonization.

\subsubsection{Real-Time Control}

In recent decades, infrastructure development, such as the deployment of smart sensors, smart meters, embedded actuators, and upgraded two-way communications, enables the real-time monitoring and control of demand-side electric devices. These advancements facilitate the implementation of demand response strategies to provide the services of frequency regulation and voltage regulation to the grid. The concept of C-VPP extends to the real-time coordinated control of demand-side DERs. However,  it is challenging to control a large population of heterogeneous DER devices due to the complexity, scalability, uncertainty, unknown models, privacy issues, and more. Advanced control techniques are necessitated to over these  obstacles and enable carbon-aware control of massive loads and DERs at scale.

\subsection{Key Issues} \label{sec:decision:issues}


Despite being discussed separately, the decarbonization decisions across the generation, grid, and demand layers are inherently interconnected. 
Achieving system-wide optimality and effectively utilizing available resources require holistic coordination among these three layers. However, the approaches for achieving such coordination are still under development, especially how to incentivize the actions and resources of vast end-users. Below we discuss several key issues in user-side decarbonization decisions.

\begin{itemize}
    \item [1)] \emph{Information Asymmetry}.
    Generally, end-users lack access to detailed power system operation information, while system operators are unaware of the specific models and statuses of private loads and DERs. Additionally, system operators generally do not have direct controllability over the private electric facilities owned by users.  \vspace{-3pt}
    \item [2)] \emph{Decentralized Decision-making}. The huge amount of user-side controllable devices poses significant scalability challenges to traditional centralized control schemes in terms of both computation and communication. Decentralized  control architectures are required to effectively  coordinate massive load devices, where each individual user or aggregator makes its own decisions and interacts with the grid.
     \vspace{-3pt}
    \item [3)] \emph{Decarbonization Targets Alignment}. Individual  users need to account for and report carbon emissions associated with their electricity consumption and are motivated to take optimal actions to reduce their own emissions. Then, a critical issue is how to \emph{align individual users' decisions with the system-level decarbonization targets}, 
so that the aggregation of individual users' actions match the optimal scheme for system-level emission reduction.  \vspace{-3pt}
    \item [4)] \emph{Smart Automated Load Management}.
   Individual users may lack the time and resources to closely monitor the grid's emission status. Achieving scalable coordination of large-scale electric facilities requires the implementation of automated and intelligent load and DER management systems. These systems, coupled with the deployment of smart meters and actuators, assist users in decision-making  and enable automatic execution. 
\end{itemize}

 The key issues above create the need for \emph{well-designed, granular, and concise control signals} from the grid that accurately reflect the actual power system emission status. Such control signals are issued to end-users to guide their decarbonization decision-making in a decentralized fashion, so that each individual user can effectively contribute to the power system decarbonization. Some potential control signals can be the real-time nodal carbon intensity, locational 
  marginal emission factors \cite{greg2022conse}, time-varying carbon pricing signal, etc.  The design of decarbonization control signals and carbon pricing schemes, along with the research on users' responses, is an emerging and critical future research direction  that requires further exploration.

%% file: Design.tex
\section{Carbon-Electricity Market Design} \label{sec:market}


With a substantial portion of the world's carbon emissions originating from power systems operating within liberalized electricity markets, any of the proposed solutions to  achieve net zero emissions necessitates significant decarbonization of the electricity sector. Incorporating carbon-related components into the formation of electricity prices can account for the externality of carbon emissions and create economic incentives for both suppliers and consumers to pursue low-carbon products and services. This transforms the existing electricity markets into \emph{carbon-electricity markets} and leads to the subject of carbon pricing. However, the market price formation is complex and requires a broad perspective that accounts for the essential roles of competition and regulation based on sound welfare economics principles. The basic principles for market price formation include: 
\begin{itemize}
    \item [1)]  \emph{Efficient Allocation}. Price signals reflect the marginal trade-off between the costs and benefits of resources and form a necessary condition for efficient allocation and attaining the maximum social gains. 
    \vspace{-3pt}
    \item [2)] \emph{Self-Revelation}. 
    It pertains to an incentive-compatible condition where every agent's optimal strategy is to act truthfully, thereby attaining the most favorable outcome by disclosing their true costs and preferences, which subsequently impact market inputs through bidding behavior.  \vspace{-3pt}
    \item [3)] \emph{Voluntary Participation}.  It warrants the freedom to make individually preferred choices, which also influences the market incentives for entry or following the system dispatch.   
    \vspace{-3pt}
    \item [4)] \emph{Revenue Sufficiency}. This refers to the requirement that market revenue needs to provide sufficient compensation to cover the costs of entering and staying in the market.
\end{itemize}
 
Electricity is a special commodity that is non-storable at scale. Within an electricity grid, the generation and load must be consistently balanced in real-time with the satisfaction of physical system constraints 
to ensure secure and stable system operation, while the power flow delivery adheres to the physical laws and network topology. This implies that the design of carbon-electricity markets should align with the physical power systems and power flows.
In the following subsections, we provide a brief overview of the electricity market, and then introduce the carbon-electricity market design and carbon pricing. Critical issues in the market design are discussed at last.

\subsection{Overview of Electricity Market}

An
electricity market typically involves four types of participants: 1) \emph{generators}  that produce electricity using  different energy sources such as coal, natural gas, wind and solar energy; 2)  \emph{transmission owners} that own and operate the transmission grids  to transport electricity from generators to distribution networks and large consumers; 3) \emph{retailers} that deliver electricity to end consumers through distribution networks; and 4) \emph{end consumers} that use electricity to power electric appliances. Accordingly,
an electricity market can be divided into two primary segments: 
\begin{itemize}
    \item [1)]  \emph{Wholesale Market} that involves the buying and selling of electricity in bulk quantities between generators, transmission owners,  retailers, and large consumers. It operates on a regional basis and is regulated by independent system operators (ISOs) or regional transmission organizations (RTOs) that manage the transmission grid and ensure the real-time power balance between
    generation and demand.
    \item[2)] \emph{Retail Market} that  involves the delivery of electricity to end consumers by retailers. The retail market is typically regulated, and the prices are set by government agencies or utility commissions, while consumers can choose from different plans or providers based on their energy needs and budget.
\end{itemize}

\subsubsection{Wholesale Electricity Market}

While wholesale electricity markets may vary in structure and procedural rules across different jurisdictions, they share the same basic elements. In particular, the primary task of power system operation involves  three key elements: 
\begin{itemize}
    \item [1)] \emph{Energy Balance}: allocation of electricity generation among suppliers and demanders that clears the market in real-time and maintains a physical balance between power generation and load demand.

    \item[2)] \emph{Transmission Congestion Management}: allocation of available transmission capacity among suppliers and demanders to maintain  system stability and mitigate forced load shedding.
        \item[3)] \emph{Reserves Provision}: provision of reserve capacities (or ancillary services) for generation and transmission to ensure power system reliability and security. 
    
\end{itemize}

Accordingly, these three elements give rise to the 
energy market, transmission market, and reserve market, respectively.
Each of these elements can be further differentiated by time scales, ranging from \emph{forward planning} to \emph{real-time operations}. 
Forward planning includes long-term resource adequacy plans, short-term day-ahead plans for the next day's operating cycle, and hour-ahead plans for the next few hours (typically 1, 2, or 4 hours).
  Real-time operations are conducted on a time frame that ranges from 1 hour down to 5 minutes prior to the actual dispatch.  
The forward markets can be viewed as financial markets, where the physical commitment of resources is indicative rather than binding, especially in long-term bilateral contracts. 
Nevertheless, as time progresses, the disparities between contracts and physical systems
diminish and are eventually 
settled at the real-time price.

The \emph{transmission market} is an extension of the energy market.  
The scarcity of transmission capacity is revealed by initial plans for electricity generation, and then adjustments in the spatial distribution of generations are made to alleviate congestion. Some power systems do not price transmission access on a spot basis, 
 but instead rely on directives from the SO that specify which generators must increase or decrease their outputs. 
 In contrast, 
 in systems that use ``\emph{congestion pricing}" of transmission, a usage charge is established based on the marginal cost of alleviating congestion. This charge can be either  on a nodal basis (e.g., PJM and CAISO in the U.S.), or on a zonal basis (e.g., western Europe countries), where only congestion between major zones is explicitly priced. 
In a zonal system, the SO uses offered bids for increments and decrements to compute the least costly way of alleviating congestion. 

In the \emph{reserves market}, 
reserves are typically provided by allocating a portion of generation capacity to a standby status or to the task of load following on a short time scale. 
The amount of reserves is typically defined as a percentage of the load, which can be divided as  primary (or ``spinning") reserves\footnote{The primary reserve usually refers to the generation capacity that can be made fully available, subject to the ramping rate limitations for thermal generators, within ten minutes and can be sustained for at least two hours. It includes upward (increasing generation) and downward (decreasing generation) reserves.}, non-synchronized (or ``non-spinning") reserves,  supplemental reserves, ``black start" reserves, etc. 
Spinning and non-spinning reserves are commonly provided by units whose bids are accepted in the energy market but do not fully exhaust their generation capacity.  
 The auction markets for reserves rely on two bid prices, one for reserved capacity and the other for the actual generation.
 


\subsubsection{Retail Electricity Market}

Distinguished from the wholesale market, the retail electricity market deals with the sale of electricity to end-consumers. The structures of retail electricity markets vary across counties. 
In some nations, a regulated rate for retail electricity is set by the government, whereas in other countries, there is competition among multiple electricity suppliers that offer a range of plans and pricing options for consumers to choose from. 
In a retail market, the electricity price is formed by a combination of factors and mechanisms. Some key factors that determine the retail electricity price include 1) the wholesale market price, 2) transmission and distribution costs, 3) system maintenance costs, 4) regulatory policies, 5) load demand, etc. In addition, retail electricity suppliers
may offer fixed-rate plans, dynamic-rate plans, or plans with time-of-use pricing, where the electricity price varies over time on a daily basis. Dynamic electricity pricing is generally designed to reflect the real-time
cost of electricity supply and guide users' energy consumption behaviors.

\subsection{Carbon Pricing and Market Mechanism}

Carbon pricing involves setting a price on the carbon content of goods and services, which is a market mechanism that internalizes the social cost of carbon emissions. In this way, carbon pricing 
creates strong incentives for both consumers and producers to engage in activities that result in reduced carbon emissions, and it is broadly recognized as one of the most efficient economic instruments available to achieve decarbonization goals at a low cost. 
According to William Nordhaus\footnote{The recipient of the 2018 Nobel Memorial Prize in Economic Sciences for integrating climate change into economic analysis.}, among many benefits of carbon pricing,  the following three are the most prominent \cite{nordhaus2013climate}: 
\begin{itemize}
    \item  Carbon pricing provides consumers with   signals about which goods and services have high carbon content, and encourages consumers to reduce their carbon footprint by  choosing low-emission options.
    
\item It provides producers with signals about which inputs lead to more carbon emissions and which ones are less or none, and thus incentivizes firms  to adopt low-carbon technologies and production processes so as to lower their costs and increase their profits.

\item  Carbon pricing generates revenue and market incentives for inventors and innovators to develop new low-carbon technologies, products, and production processes,  thus accelerating the transition to a low-carbon economy.
\end{itemize}


\subsubsection{Carbon Pricing Mechanisms}

There are a variety of carbon pricing mechanisms available, and governments and businesses can determine the most appropriate schemes based on their specific policy environment and objectives. 
The World Bank Group and the Organization for Economic Co-operation and Development (OECD) have developed general guidelines for effective carbon pricing schemes, known as the ``FASTER Principles" \cite{faster}: fairness, alignment of policy and objectives, stability and predictability, transparency, efficiency and cost-effectiveness, and reliability and environmental integrity. Below,
we introduce two examples of carbon pricing mechanisms in the electricity market:
\begin{itemize}
    \item [1)] \emph{Direct Carbon Pricing}  
    places a volumetric cost adder on electricity generation, proportional to the amount of carbon emissions. As a result, the market price reflects the externality cost of carbon emissions, and consumers pay for every ton of carbon pollution resulting from their activities. 
    Such a mechanism creates a financial incentive to reduce emissions by switching to more efficient processes or cleaner fuels. In addition,
    this approach offers a high degree of price certainty, given that the price per ton of pollution is relatively fixed, although the extent of emissions reduction remains uncertain.

  \item [2)] \emph{Cap-and-Trade Mechanism}  sets a limit on total carbon emissions from specific regions or entities, and issues carbon permits or allowances so that the ``carbon emission rights" can be traded in a market,  thereby establishing carbon prices. This  forms an emission trading system and allows polluters the flexibility to meet emissions reduction targets at a low cost. It   provides certainty about emission reductions, while the price for carbon emission rights remains uncertain.

\end{itemize}

Carbon pricing schemes operate most effectively when not hindered by conflicting policy interventions. Additionally, to achieve efficient price formation results, a rigorous carbon accounting system based on sound principles is essential and foundational. Such a carbon accounting system provides the market mechanisms with an unambiguous determination of the carbon content of goods and services, with precise attributions of the sources of carbon emissions. We refer readers to Section \ref{sec:accounting} of this paper
for an in-depth overview and discussion of carbon accounting.

\subsubsection{Potential Carbon-Electricity Pricing Schemes}

The direct carbon pricing mechanism introduced above can be adopted in the electricity markets, leading to the establishment of \emph{carbon-aware electricity pricing}. 
The benefit of doing so is that  existing electricity pricing mechanisms can still be utilized with minor adjustments. Below
we propose specific carbon-aware electricity pricing approaches in the electricity market:
\begin{itemize}
    \item In day-ahead wholesale markets,  the electricity price is typically determined by an auction-bidding process, where the cost of carbon emissions needs to be incorporated into the generation cost for bidding.

    \item   In real-time wholesale markets, the locational marginal price (LMP) is primarily used for electricity pricing.  The LMPs are typically calculated by solving an optimal power flow (OPF)-based economic dispatch problem, and the LMPs are determined as the dual variables (known as the ``shadow prices") of real power balance constraints of the OPF problem \cite{orfanogianni2007general}. Similarly, one can solve the carbon-aware optimal power flow (C-OPF) problem presented in Section \ref{sec:copf} and take the corresponding shadow prices as the real-time nodal electricity prices, termed as
    Carbon-aware LMP (C-LMP).

    \item In the retail markets, dynamic price signals have been employed to guide end-consumers to adjust their electricity usage to fit the grid's needs, which is also referred to as price-based demand response. Commonly-used dynamic pricing schemes include real-time pricing (RTP), time-of-use (TOU) pricing, critical peak pricing (CPP), etc. \cite{yan2018review} As aforementioned in Section \ref{sec:decision:issues}, the dynamic pricing signal can serve as a promising control signal to inform user-side decarbonization decision-making. To this end, one can design carbon-aware dynamic pricing mechanisms, such as the carbon-aware versions of RTP, TOU, and CPP, to incorporate the time-varying grid emission status into the electricity price.  
    Intuitively, when the grid is in a high-emission state or the nodal carbon intensity is higher, the price tends to be higher as well. This is intended to discourage users' electricity usage during such periods.
    Conversely, when the grid is in low emissions, the price tends to be lower, encouraging users to shift their electricity consumption to this period.
    
\end{itemize}

An alternative is to apply the cap-and-trade mechanism for carbon pricing. In this way, carbon emissions are treated as a commodity with their quantity determined by carbon accounting and the price  determined by the carbon trading market. The advantage is that the carbon market and carbon pricing can be independent of the electricity market, and there is no need to alter the existing electricity market structure. The employment of flow-based carbon accounting can align the determination of user-side emission quantities with the physical electricity  supply and use. In \cite{hua2020blockchain}, a blockchain-based peer-to-peer trading framework is proposed to trade energy and carbon allowance.

\subsubsection{Current Industry Practice}

\begin{figure}
    \centering
    \includegraphics[scale=0.45]{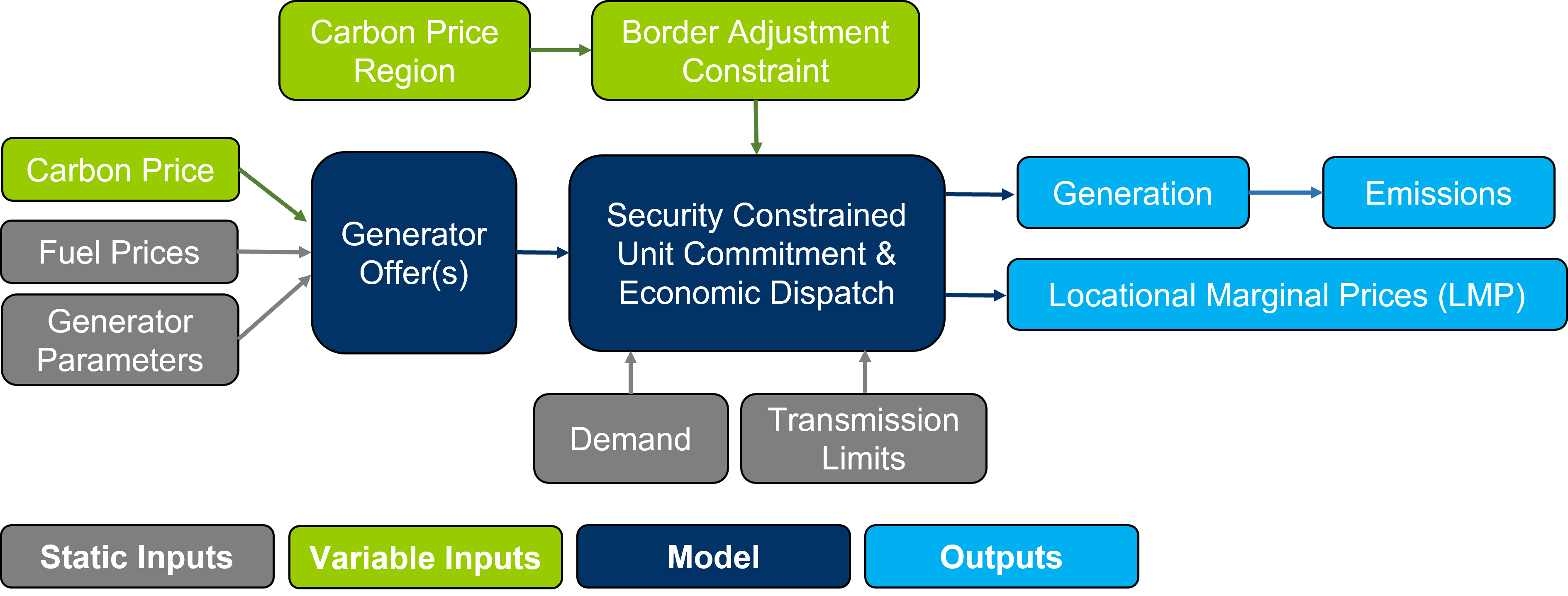}
    \caption{Modeling framework for carbon pricing at PJM  \cite{pjmcarbonp}.    }
    \label{fig:modelfr}
\end{figure}

The
carbon-electricity pricing has attracted a great deal of recent attention from the industry. A number of 
ISOs, utilities, and federal agencies have conducted studies on carbon pricing in electricity markets, such as 
CAISO \cite{CAISOcp}, ISO New England \cite{ISONcp},  New York ISO \cite{NYSOcp}, PJM \cite{pjmcarbonp}, Federal Energy Regulatory Commission of Department of Energy \cite{DOEcp}, and SCE Dynamic Rate Pilot program \cite{scecp}. We use PJM as a case study to exemplify the practical design of carbon pricing. In this context, it is assumed that the carbon price is determined by policymakers external to the market. As shown in Figure \ref{fig:modelfr}, at the core of the  PJM  market management system is a market clearing process that relies on an optimal power flow model, also known as the market clearing engine (MCE), to perform security-constrained unit commitment and economic dispatch. By far,
generator offers are the most significant inputs into the market-clearing process, while price-responsive demand bids are also crucial but have a much smaller volume.  Transmission limits are another important set of input parameters that are used for congestion management and the determination of nodal prices. However, it is challenging to implement carbon pricing schemes in practice due to a variety of factors, including the resistance from multiple stakeholders towards the adoption of new pricing mechanisms. Other key issues on the design and implementation of carbon pricing are discussed in the next subsection.


\subsection{Key Issues}


An important takeaway is that the research advancement is of fundamental importance to provide a solid foundation for future market-based decarbonization solutions. 
Looking ahead, we identify several key issues that offer research opportunities for electricity market design, including the following:

\begin{itemize}
    \item \emph{Carbon Border Adjustment}. In practice, when carbon pricing policies and regulations are inconsistent at regional or global levels, 
an important problem called “\emph{carbon leakage}” may occur. This means that carbon-intensive industries or firms shift their operations or activities  from jurisdictions with strict carbon pricing policies (or high carbon cost) to those with less stringent policies (or lower carbon cost). 
As a consequence of the shifts in economic activities caused by the phenomenon of carbon leakage, carbon pricing may result in higher costs that are undesirably disproportionate to the reduction in total carbon emissions achieved. Therefore, 
 \emph{carbon border adjustment} is designed to  avert carbon leakage and ensure that carbon pricing policies remain effective in reducing carbon emissions.  It also 
  protects domestic industrial competitiveness by reducing the incentive for firms to move production abroad.  Essentially, carbon border adjustment can be thought of as a combination of regional ``\emph{carbon taxes}" for imports and ``\emph{carbon rebates}" for exports.

   \item \emph{Demand-Side Carbon Pricing}.  Most existing carbon pricing schemes focus on the pricing of generation-side carbon emissions. However, the design of demand-side carbon pricing or other incentive mechanisms is also necessary to fully exploit the power flexibility of vast end-users for decarbonization. 
A critical question is how to design markets to reflect the value of demand-side flexible resources, such as coordinated electric vehicle charging, energy storage systems, load management of space/water heaters and  heating,
ventilation, and air conditioning (HVAC) systems, and to incentivize their operation for system-wise decarbonization.

\item \emph{Testing and Evaluation}. Comprehensive research and analytical tools are required to  rigorously test and evaluate various market design proposals. This entails developing game theoretic models of electricity generator behavior and market clearing engines as foundational elements for such  work. A ``market design laboratory” is needed to test the effectiveness of various designs in terms of investment requirements, spot price behavior, carbon emissions levels, etc. Additionally, establishing comprehensive databases, standardized testing processes, and authentic carbon accounting frameworks, is essential to provide policy-makers and market designers with an evidence-based foundation to guide and advise the process of market-driven decarbonization. It is also important to 
study the interaction and coordination among different parts of the market, including wholesale, retail, and capacity markets, as well as regulatory policies.
 
\end{itemize}

There are 
numerous specific questions arising in terms of the future market design, such as 1) how to design practically applicable  carbon pricing or cap-and-trade mechanisms to incentivize carbon emission reduction and promote innovations and investments in clean energy technologies? 2) what type of market design could accommodate a high penetration of intermittent renewable generation and incentivize energy storage technologies? 3) how to advance the development of long-term contracts and forward markets to lower the risk and cost to achieve carbon neutrality? and more. Essentially, decarbonization is a public good, and thus policy and regulation are necessitated to address the fundamental issues of standardization, infrastructure, fairness, and equity. 

%% file: Conclusion.tex
 \section{Conclusion and Future Vision}\label{sec:conclusion}



This paper presents a comprehensive framework that addresses pivotal challenges of carbon-electricity nexus in electric power systems, including carbon accounting, carbon-aware decision-making, and carbon-electricity market design. As a distinguishing feature of this framework, the flow-based nature aligns carbon emissions with the physical power grid and power flow and leads to high temporal and spatial granularity. 
The proposed framework can serve as a 
comprehensive guide for future research, education programs, policy development, and business initiatives concerning decarbonization in the electric power
sector. By providing effective strategies and solutions, the proposed framework supports the overarching goal of decarbonization and addresses the urgent need to combat climate change.  Below, we close this paper by discussing the key challenges and future vision in the implementation of this flow-based framework and other granular carbon-aware approaches.

  \vspace{3pt}
 \noindent
$\bullet$ \textbf{Data Availability, Quality, and Integration}. 
As previously mentioned, the availability of high-quality up-to-date granular data is crucial for accurate carbon accounting and optimal decarbonization decision-making. It requires power plants, ISOs, and utilities to disclose the system's carbon emission status data in a timely way. Transparent data-sharing practices and standardized reporting frameworks help enhance the credibility and reliability of the data. In addition, ensuring data quality is important, as practical system data often encounter a variety of issues, such as missing
data, outlier data, noisy data, and outdated information, and thus pre-processing and sanity check on
raw data is needed. 
Integrating data from various sources and formats, including power systems, grid operations, user-side devices, and emission monitoring systems, also presents challenges. Harmonizing and combining diverse datasets to obtain a comprehensive understanding of carbon emissions is vital for informing decision-making and market design.
 To address these data issues, 
authentic data management platforms, advanced data analytics techniques, data sharing agreements, and collaboration among different stakeholders, are necessitated.

\vspace{3pt}
\noindent
$\bullet$ \textbf{Infrastructure Development}. On the generation side, continuous emission monitoring systems,  gas chromatographs, mass flow meters, or other emission measurement devices need to be deployed to continuously monitor and measure carbon emissions, which 
 provide real-time accurate  data on physical emissions from power generators. On the demand side, the two-way communication infrastructure between power grids and end-users is necessary to enable the transmission of important emission information and control signals. Moreover, the deployment of advanced metering infrastructures (AMI)
allows users to stay informed about the system's emission status. Smart meters, for instance, not only can display the real-time electricity price and energy consumption but also provide information on nodal carbon intensity or dynamic carbon prices, enabling users to monitor and respond to the carbon impact of their electricity consumption. In addition, embedded controllers, smart actuators, distributed computing units, and intelligent load management systems are needed to facilitate the automation and implementation of user-side carbon-aware activities.

\vspace{3pt}
\noindent
$\bullet$ \textbf{Standardization}. Standardization plays a crucial role in ensuring the comparability, compatibility, and transparency of carbon emissions data and infrastructure.  
Standardizing carbon accounting methodologies, evaluation processes, calculations, and data sharing is essential for consistent and accurate reporting and tracking of emissions across various entities. This fosters fair and effective comparisons of emissions and builds trust in the carbon market. Moreover, the development of standardized modeling tools and simulation platforms is required to test the impact of different energy policies, regulations, and market mechanisms on carbon emissions reduction and overall system performance. The electricity sector is one of the most complex societal-cyber-physical engineering systems, and thus standardization needs collaboration among all stakeholders to promote coordinated and effective decarbonization. 


\vspace{3pt}
\noindent
$\bullet$ \textbf{Fairness and Equity}. The importance of fairness in carbon accounting has been discussed in Section \ref{sec:discussion}. In addition, it is essential to consider the fairness issues pertaining to carbon-related data, decision-making, policy development, and market design.
These activities can have uneven impacts on different segments of society, with some groups experiencing disproportionate benefits or burdens. For instance, decarbonization initiatives and policies may result in higher electricity prices, which can largely affect low-income communities and raise concerns about energy affordability and access to basic energy services. Furthermore, certain communities or individuals may have limited access to emission information, low-carbon technology, and other necessary resources. 
 This can lead to unequal outcomes from carbon-neural energy solutions and further exacerbate social and economic disparities.
Addressing these fairness and equity issues is critical to ensuring that carbon accounting, carbon-aware decision-making, and market operation in power systems contribute to a more just and sustainable energy future for all.


%% file: main.bbl
\begin{thebibliography}{10}

\bibitem{portner2022climate}
Hans-O P{\"o}rtner, Debra~C Roberts, Helen Adams, Carolina Adler, Paulina Aldunce, Elham Ali, Rawshan~Ara Begum, Richard Betts, Rachel~Bezner Kerr, Robbert Biesbroek, et~al.
\newblock {\em Climate change 2022: Impacts, adaptation and vulnerability}.
\newblock IPCC Geneva, Switzerland, 2022.

\bibitem{carbonclock}
{World Emission Clock}.
\newblock 2023.
\newblock \url{https://worldemissions.io/}.

\bibitem{google247program}
Google.
\newblock Moving toward 24$\times$7 carbon-free energy at google data centers: Progress and insights.
\newblock 2018.
\newblock \url{https://www.gstatic.com/gumdrop/sustainability/24x7-carbon-free-energy-data-centers.pdf}.

\bibitem{microsoft}
Microsoft.
\newblock Expanding carbon-free electricity globally: Microsoft electricity policy brief.
\newblock 2022.
\newblock \url{https://query.prod.cms.rt.microsoft.com/cms/api/am/binary/RE57d2R}.

\bibitem{brander2018creative}
Matthew Brander, Michael Gillenwater, and Francisco Ascui.
\newblock Creative accounting: A critical perspective on the market-based method for reporting purchased electricity (scope 2) emissions.
\newblock {\em Energy Policy}, 112:29--33, 2018.

\bibitem{bjorn2022rec}
Anders Bjorn, Shannon Lloyd, Matthew Brander, and Damon Matthews.
\newblock Renewable energy certificates allow companies to overstate their emission reductions.
\newblock {\em Nature Climate Change}, 2022.

\bibitem{world2004greenhouse}
Janet Ranganathan, Laurent Corbier, Pankaj Bhatia, et~al.
\newblock The greenhouse gas protocol: A corporate accounting and reporting standard, 2004.

\bibitem{world2014scope2}
Mary Sotos.
\newblock Ghg protocol scope 2 guidance, 2015.

\bibitem{giannakis2013monitoring}
Georgios~B Giannakis, Vassilis Kekatos, Nikolaos Gatsis, Seung-Jun Kim, Hao Zhu, and Bruce~F Wollenberg.
\newblock Monitoring and optimization for power grids: A signal processing perspective.
\newblock {\em IEEE Signal Processing Magazine}, 30(5):107--128, 2013.

\bibitem{grainger1999power}
John~J Grainger.
\newblock {\em Power system analysis}.
\newblock McGraw-Hill, 1999.

\bibitem{miller2022hourly}
Gregory~J Miller, Kevin Novan, and Alan Jenn.
\newblock Hourly accounting of carbon emissions from electricity consumption.
\newblock {\em Environmental Research Letters}, 17(4):044073, 2022.

\bibitem{eGRID}
{US Environmental Protection Agency (EPA) eGRID}.
\newblock 2023.
\newblock \url{https://www.epa.gov/egrid}.

\bibitem{eia}
{US Energy Informaton Administration}.
\newblock 2023.
\newblock \url{https://www.epa.gov/egrid}.

\bibitem{isone}
{ISO New England}.
\newblock 2023.
\newblock \url{https://www.iso-ne.com/system-planning/system-plans-studies/emissions}.

\bibitem{CAISO}
{California Independent System Operator}.
\newblock 2023.
\newblock \url{http://www.caiso.com/todaysoutlook/pages/emissions.html}.

\bibitem{singu}
{Open Grid Emissions from Singularity Energy}.
\newblock 2023.
\newblock \url{https://singularity.energy/open-grid-emissions/}.

\bibitem{de2019tracking}
Jacques~A de~Chalendar, John Taggart, and Sally~M Benson.
\newblock Tracking emissions in the us electricity system.
\newblock {\em Proceedings of the National Academy of Sciences}, 116(51):25497--25502, 2019.

\bibitem{lau2008bottom}
Chris Lau and Jaineel Aga.
\newblock Bottom line on renewable energy certificates. {[Online]: \url{https://www.wri.org/research/bottom-line-renewable-energy-certificates}.}
\newblock 2008.

\bibitem{jones2015legal}
Todd Jones, Robin Quarrier, Maya Kelty, et~al.
\newblock The legal basis for renewable energy certificates.
\newblock {\em Center for Resource Solutions. Retrieved May}, 25:2021, 2015.

\bibitem{kansal2018introduction}
Rachit Kansal.
\newblock Introduction to the virtual power purchase agreement.
\newblock {\em Rocky Mountain Institute}, 2018.

\bibitem{tang2019classification}
Chenghui Tang and Fan Zhang.
\newblock Classification, principle and pricing manner of renewable power purchase agreement.
\newblock In {\em IOP Conference Series: Earth and Environmental Science}, volume 295, page 052054. IOP Publishing, 2019.

\bibitem{nrelgreen}
Jenny Heeter, Eric O’Shaughnessy, and Rebecca Burd.
\newblock Status and trends in the voluntary market (2020 data), 2021.

\bibitem{247comp}
{United Nations: 24/7 Carbon-free Energy Compact}.
\newblock 2023.
\newblock \url{https://www.un.org/en/energy-compacts/page/compact-247-carbon-free-energy}.

\bibitem{xu2472021}
Qingyu Xu, Aneesha Manocha, Neha Patankar, and Jesse Jenkins.
\newblock System-level impacts of 24/7 carbon-free electricity procurement.
\newblock {\em Available at SSRN 4248431}, 2021.

\bibitem{miller2020beyond}
Gregory Miller.
\newblock Beyond 100\% renewable: Policy and practical pathways to 24/7 renewable energy procurement.
\newblock {\em The Electricity Journal}, 33(2):106695, 2020.

\bibitem{mic247}
Alix~de Monts, Diego~Hernandez Diaz, Florian Kühn, and {et al.}
\newblock A path towards full grid decarbonization with 24/7 clean power purchase agreements. {[Online]: \url{https://www.mckinsey.com/industries/electric-power-and-natural-gas/our-insights/decarbonizing-the-grid-with-24-7-clean-power-purchase-agreements}}, 2021.

\bibitem{247epri2022}
Adam Diamant.
\newblock 24/7 carbon-free energy: Matching carbon-free energy procurement to hourly electric load, 2022.

\bibitem{bjorn2022renewable}
Anders Bjorn, Shannon Lloyd, Matthew Brander, and Damon Matthews.
\newblock Renewable energy certificates threaten the integrity of corporate science-based targets.
\newblock {\em Nature Climate Change}, 2022.

\bibitem{monyei2018electrons}
Chukwuka~G Monyei and Kirsten~EH Jenkins.
\newblock Electrons have no identity: Setting right misrepresentations in google and apple’s clean energy purchasing.
\newblock {\em Energy research \& social science}, 46:48--51, 2018.

\bibitem{ricks2023minimizing}
Wilson Ricks, Qingyu Xu, and Jesse~D Jenkins.
\newblock Minimizing emissions from grid-based hydrogen production in the united states.
\newblock {\em Environmental Research Letters}, 18(1):014025, 2023.

\bibitem{resi2021}
{CRS}.
\newblock Data sources: Accounting for standard delivery renewable energy. {[Online]: \url{https://resource-solutions.org/wp-content/uploads/2021/03/Data-Sources-Standard-Delivery.pdf}}, 2021.

\bibitem{greene}
{Green-e}.
\newblock 2023.
\newblock \url{https://www.green-e.org/residual-mix\#_ftn4}.

\bibitem{broekhoff2007guidelines}
Derik Broekhoff.
\newblock Guidelines for quantifying ghg reductions from grid-connected electricity projects.
\newblock 2007.

\bibitem{greg2022conse}
Gregory Miller.
\newblock Applying the consequential emissions framework for emissions-optimized decision-making for energy procurement and management.
\newblock 2022.

\bibitem{siler2012marginal}
Kyle Siler-Evans, In{\^e}s~Lima Azevedo, and M~Granger Morgan.
\newblock Marginal emissions factors for the us electricity system.
\newblock {\em Environmental science \& technology}, 46(9):4742--4748, 2012.

\bibitem{rothschild2009total}
Susy Rothschild and Art Diem.
\newblock Total, non-baseload, {eGRID} subregion, state? guidance on the use of {eGRID} output emission rates.
\newblock In {\em 18th Annual International Emission Inventory Conference" Comprehensive Inventories-Leveraging Technology and Resources". Baltimore, MD}, 2009.

\bibitem{zheng2015assessment}
Zhanghua Zheng, Fengxia Han, Furong Li, and Jiahui Zhu.
\newblock Assessment of marginal emissions factor in power systems under ramp-rate constraints.
\newblock {\em CSEE Journal of Power and Energy Systems}, 1(4):37--49, 2015.

\bibitem{wang2014locational}
Y~Wang, C~Wang, CJ~Miller, SP~McElmurry, SS~Miller, and MM~Rogers.
\newblock Locational marginal emissions: Analysis of pollutant emission reduction through spatial management of load distribution.
\newblock {\em Applied energy}, 119:141--150, 2014.

\bibitem{deetjen2019reduced}
Thomas~A Deetjen and In{\^e}s~L Azevedo.
\newblock Reduced-order dispatch model for simulating marginal emissions factors for the united states power sector.
\newblock {\em Environmental science \& technology}, 53(17):10506--10513, 2019.

\bibitem{baumgartner2019design}
Nils Baumg{\"a}rtner, Roman Delorme, Maike Hennen, and Andr{\'e} Bardow.
\newblock Design of low-carbon utility systems: Exploiting time-dependent grid emissions for climate-friendly demand-side management.
\newblock {\em Applied energy}, 247:755--765, 2019.

\bibitem{rogers2013evaluation}
Michelle~M Rogers, Yang Wang, Caisheng Wang, Shawn~P McElmurry, and Carol~J Miller.
\newblock Evaluation of a rapid lmp-based approach for calculating marginal unit emissions.
\newblock {\em Applied energy}, 111:812--820, 2013.

\bibitem{wang2016estimating}
Caisheng Wang, Yang Wang, Carol~J Miller, and Jeremy Lin.
\newblock Estimating hourly marginal emission in real time for pjm market area using a machine learning approach.
\newblock In {\em 2016 IEEE Power and Energy Society General Meeting (PESGM)}, pages 1--5. IEEE, 2016.

\bibitem{corradi2018estimating}
Olivier Corradi.
\newblock Estimating the marginal carbon intensity of electricity with machine learning.
\newblock {\em ElectricityMap nd https://electricitymap. org/blog/marginal-carbon-in t ensity-of-electricity-with-machine-learning/.[Accessed 10 December 2021]. accessed}, 2018.

\bibitem{wattime2022}
{WattTime}.
\newblock Marginal emissions modeling: Watttime’s approach to modeling and validation.
\newblock 2022.
\newblock \url{https://www.watttime.org/app/uploads/2022/10/WattTime-MOER-modeling-20221004.pdf}.

\bibitem{greg2022marginal}
Gregory Miller.
\newblock Guide to sourcing marginal emissions factor data.
\newblock 2022.
\newblock \url{https://cebi.org/wp-content/uploads/2022/11/Guide-to-Sourcing-Marginal-Emissions-Factor-Data.pdf}.

\bibitem{mefpjm}
{PJM}.
\newblock Marginal emissions rate -- a primer.
\newblock 2022.
\newblock \url{https://pjm.com/-/media/etools/data-miner-2/marginal-emissions-primer.ashx}.

\bibitem{lmeresurety}
David~Luke Oates and Kathleen Spees.
\newblock Locational marginal emissions: A force multiplier for the carbon impact of clean energy programs, 2022.

\bibitem{atkinson2011trade}
Giles Atkinson, Kirk Hamilton, Giovanni Ruta, and Dominique Van Der~Mensbrugghe.
\newblock Trade in ‘virtual carbon’: Empirical results and implications for policy.
\newblock {\em Global Environmental Change}, 21(2):563--574, 2011.

\bibitem{iniCEF}
Anonymous.
\newblock Carbon flows: The emissions omitted: The usual figures ignore the role of trade in the world’s carbon economy.
\newblock Economist, 2011.
\newblock \url{http://www.economist.com/node/18618451}.

\bibitem{kang2012carbon}
Chongqing Kang, Tianrui Zhou, Qixin Chen, Qianyao Xu, Qing Xia, and Zhen Ji.
\newblock Carbon emission flow in networks.
\newblock {\em Scientific reports}, 2(1):479, 2012.

\bibitem{kang2015carbon}
Chongqing Kang, Tianrui Zhou, Qixin Chen, Jianhui Wang, Yanlong Sun, Qing Xia, and Huaguang Yan.
\newblock Carbon emission flow from generation to demand: A network-based model.
\newblock {\em IEEE Transactions on Smart Grid}, 6(5):2386--2394, 2015.

\bibitem{chen2019tracing}
Yu~Christine Chen and Sairaj~V Dhople.
\newblock Tracing power with circuit theory.
\newblock {\em IEEE Transactions on Smart Grid}, 11(1):138--147, 2019.

\bibitem{chang2001electricity}
Ya-Chin Chang and Chan-Nan Lu.
\newblock An electricity tracing method with application to power loss allocation.
\newblock {\em International Journal of Electrical Power \& Energy Systems}, 23(1):13--17, 2001.

\bibitem{bialek1996tracing}
Janusz Bialek.
\newblock Tracing the flow of electricity.
\newblock {\em IEE Proceedings-Generation, Transmission and Distribution}, 143(4):313--320, 1996.

\bibitem{chen2023carbon}
Xin Chen, Andy Sun, Wenbo Shi, and Na~Li.
\newblock Carbon-aware optimal power flow.
\newblock {\em arXiv preprint arXiv:2308.03240}, 2023.

\bibitem{chao2000flow}
Hung-po Chao, Stephen Peck, Shmuel Oren, and Robert Wilson.
\newblock Flow-based transmission rights and congestion management.
\newblock {\em The Electricity Journal}, 13(8):38--58, 2000.

\bibitem{totalCA2021}
{Kevala}.
\newblock Total carbon accounting: A framework to deliver locational carbon intensity data. {[Online]: \url{https://kevala.com/wp-content/uploads/2021/11/Total-Carbon-Accounting.pdf}}, 2021.

\bibitem{SCEpathway}
{Southern California Edison}.
\newblock {PATHWAY} 2045: Update to the clean power and electrification pathway.
\newblock 2019.
\newblock \url{https://www.edison.com/our-perspective/pathway-2045}.

\bibitem{xie2021toward}
Le~Xie, Chanan Singh, Sanjoy~K Mitter, Munther~A Dahleh, and Shmuel~S Oren.
\newblock Toward carbon-neutral electricity and mobility: Is the grid infrastructure ready?
\newblock {\em Joule}, 5(8):1908--1913, 2021.

\bibitem{frank2012optimal}
Stephen Frank, Ingrida Steponavice, and Steffen Rebennack.
\newblock Optimal power flow: a bibliographic survey i.
\newblock {\em Energy systems}, 3(3):221--258, 2012.

\bibitem{cain2012history}
Mary~B Cain, Richard~P O’neill, Anya Castillo, et~al.
\newblock History of optimal power flow and formulations.
\newblock {\em Federal Energy Regulatory Commission}, 1:1--36, 2012.

\bibitem{cheng2018bi}
Yaohua Cheng, Ning Zhang, and Chongqing Kang.
\newblock Bi-level expansion planning of multiple energy systems under carbon emission constraints.
\newblock In {\em 2018 IEEE Power \& Energy Society General Meeting (PESGM)}, pages 1--5. IEEE, 2018.

\bibitem{wei2021carbon}
Xiang Wei, Xian Zhang, Yuxin Sun, and Jing Qiu.
\newblock Carbon emission flow oriented tri-level planning of integrated electricity--hydrogen--gas system with hydrogen vehicles.
\newblock {\em IEEE Transactions on Industry Applications}, 58(2):2607--2618, 2021.

\bibitem{wu2022carbon}
Ting Wu, Xiang Wei, Xian Zhang, Guibin Wang, Jing Qiu, and Shiwei Xia.
\newblock Carbon-oriented expansion planning of integrated electricity-natural gas systems with ev fast-charging stations.
\newblock {\em IEEE Transactions on Transportation Electrification}, 8(2):2797--2809, 2022.

\bibitem{sun2017analysis}
Yanlong Sun, Chongqing Kang, Qing Xia, Qixin Chen, Ning Zhang, and Yaohua Cheng.
\newblock Analysis of transmission expansion planning considering consumption-based carbon emission accounting.
\newblock {\em Applied energy}, 193:232--242, 2017.

\bibitem{viral2012optimal}
Rajkumar Viral and Dheeraj~Kumar Khatod.
\newblock Optimal planning of distributed generation systems in distribution system: A review.
\newblock {\em Renewable and sustainable energy Reviews}, 16(7):5146--5165, 2012.

\bibitem{shen2020low}
Wei Shen, Jing Qiu, Ke~Meng, Xi~Chen, and Zhao~Yang Dong.
\newblock Low-carbon electricity network transition considering retirement of aging coal generators.
\newblock {\em IEEE Transactions on Power Systems}, 35(6):4193--4205, 2020.

\bibitem{tao2020carbon}
Yuechuan Tao, Jing Qiu, Shuying Lai, Junhua Zhao, and Yusheng Xue.
\newblock Carbon-oriented electricity network planning and transformation.
\newblock {\em IEEE Transactions on Power Systems}, 36(2):1034--1048, 2020.

\bibitem{padhy2004unit}
Narayana~Prasad Padhy.
\newblock Unit commitment-a bibliographical survey.
\newblock {\em IEEE Transactions on power systems}, 19(2):1196--1205, 2004.

\bibitem{xia2010optimal}
X~Xia and AM~Elaiw.
\newblock Optimal dynamic economic dispatch of generation: A review.
\newblock {\em Electric power systems research}, 80(8):975--986, 2010.

\bibitem{zhang2020advances}
Zhien Zhang, Tao Wang, Martin~J Blunt, Edward~John Anthony, Ah-Hyung~Alissa Park, Robin~W Hughes, Paul~A Webley, and Jinyue Yan.
\newblock Advances in carbon capture, utilization and storage, 2020.

\bibitem{bevrani2017intelligent}
Hassan Bevrani and Takashi Hiyama.
\newblock {\em Intelligent automatic generation control}.
\newblock CRC press, 2017.

\bibitem{rosso2021grid}
Roberto Rosso, Xiongfei Wang, Marco Liserre, Xiaonan Lu, and Soenke Engelken.
\newblock Grid-forming converters: Control approaches, grid-synchronization, and future trends—a review.
\newblock {\em IEEE Open Journal of Industry Applications}, 2:93--109, 2021.

\bibitem{pattabiraman2018comparison}
Dinesh Pattabiraman, RH~Lasseter, and TM~Jahns.
\newblock Comparison of grid following and grid forming control for a high inverter penetration power system.
\newblock In {\em 2018 IEEE Power \& Energy Society General Meeting (PESGM)}, pages 1--5. IEEE, 2018.

\bibitem{chen2010power}
Qixin Chen, Chongqing Kang, Qing Xia, and Jin Zhong.
\newblock Power generation expansion planning model towards low-carbon economy and its application in china.
\newblock {\em IEEE Transactions on Power Systems}, 25(2):1117--1125, 2010.

\bibitem{shao2019low}
Changzheng Shao, Yi~Ding, and Jianhui Wang.
\newblock A low-carbon economic dispatch model incorporated with consumption-side emission penalty scheme.
\newblock {\em Applied Energy}, 238:1084--1092, 2019.

\bibitem{wang2021optimal}
Yunqi Wang, Jing Qiu, and Yuechuan Tao.
\newblock Optimal power scheduling using data-driven carbon emission flow modelling for carbon intensity control.
\newblock {\em IEEE Transactions on Power Systems}, 2021.

\bibitem{esaccount2021}
{Washington Department of Commerce}.
\newblock Energy storage accounting issues, 2021.

\bibitem{esacc2022}
{Federal Energy Regulatory Commission}.
\newblock Accounting and reporting treatment of certain renewable energy assets, 2022.

\bibitem{chen2021model}
Xin Chen, Jorge~I Poveda, and Na~Li.
\newblock Model-free optimal voltage control via continuous-time zeroth-order methods.
\newblock {\em arXiv preprint arXiv:2103.14703}, 2021.

\bibitem{sun2019review}
Hongbin Sun, Qinglai Guo, Junjian Qi, Venkataramana Ajjarapu, Richard Bravo, Joe Chow, Zhengshuo Li, Rohit Moghe, Ehsan Nasr-Azadani, Ujjwol Tamrakar, et~al.
\newblock Review of challenges and research opportunities for voltage control in smart grids.
\newblock {\em IEEE Transactions on Power Systems}, 34(4):2790--2801, 2019.

\bibitem{6298199}
Chuangang Ren, Di~Wang, Bhuvan Urgaonkar, and Anand Sivasubramaniam.
\newblock Carbon-aware energy capacity planning for datacenters.
\newblock In {\em 2012 IEEE 20th International Symposium on Modeling, Analysis and Simulation of Computer and Telecommunication Systems}, pages 391--400, 2012.

\bibitem{pepper2022our}
Jan Pepper, Greg Miller, Siobhan Doherty, Sara Maatta, and Mehdi Shahriari.
\newblock Our path to 24/7 renewable energy by 2025.
\newblock In {\em Proceedings of the American Solar Energy Society National Conference: ASES SOLAR 2022}, pages 36--48. Springer, 2022.

\bibitem{siano2014demand}
Pierluigi Siano.
\newblock Demand response and smart grids—a survey.
\newblock {\em Renewable and sustainable energy reviews}, 30:461--478, 2014.

\bibitem{chen2021online}
Xin Chen, Yingying Li, Jun Shimada, and Na~Li.
\newblock Online learning and distributed control for residential demand response.
\newblock {\em IEEE Transactions on Smart Grid}, 12(6):4843--4853, 2021.

\bibitem{nordhaus2013climate}
William Nordhaus.
\newblock {\em The climate casino: Risk, uncertainty, and economics for a warming world}.
\newblock Yale University Press, 2013.

\bibitem{faster}
{OECD and World Bank Group}.
\newblock The {FASTER} principles for successful carbon pricing: An approach based on initial experience.
\newblock 2015.
\newblock \url{https://www.oecd.org/environment/tools-evaluation/FASTER-carbon-pricing.pdf}.

\bibitem{orfanogianni2007general}
Tina Orfanogianni and George Gross.
\newblock A general formulation for lmp evaluation.
\newblock {\em IEEE Transactions on Power Systems}, 22(3):1163--1173, 2007.

\bibitem{yan2018review}
Xing Yan, Yusuf Ozturk, Zechun Hu, and Yonghua Song.
\newblock A review on price-driven residential demand response.
\newblock {\em Renewable and Sustainable Energy Reviews}, 96:411--419, 2018.

\bibitem{hua2020blockchain}
Weiqi Hua, Jing Jiang, Hongjian Sun, and Jianzhong Wu.
\newblock A blockchain based peer-to-peer trading framework integrating energy and carbon markets.
\newblock {\em Applied Energy}, 279:115539, 2020.

\bibitem{CAISOcp}
Danny Cullenward.
\newblock Carbon pricing in california, caiso, \& the eim.
\newblock 2020.
\newblock \url{https://policyintegrity.org/documents/Danny_Cullenward\%2C_Stanford_University.pdf}.

\bibitem{ISONcp}
Joseph Cavicchi.
\newblock Carbon pricing for new england.
\newblock 2020.
\newblock \url{https://www.iso-ne.com/static-assets/documents/2020/09/2020_09_17_iso_ne_clg_carbon_pricing_for_ne_cavicchi.pdf}.

\bibitem{NYSOcp}
Ethan~D. Avallone.
\newblock Carbon pricing: Market design complete.
\newblock 2019.
\newblock \url{https://www.nyiso.com/documents/20142/7129597/6.20.2019_MIWG_Carbon_Pricing_MDC_FINAL.pdf/cf67ebb8-d0fc-7b4b-100f-c3756d6afae8}.

\bibitem{pjmcarbonp}
{PJM Carbon Pricing Senior Task Force}.
\newblock Expanded results of {PJM} study of carbon pricing \& potential leakage mitigation mechanisms.
\newblock 2020.
\newblock \url{https://www.pjm.com/-/media/committees-groups/task-forces/cpstf/2020/20200225/20200225-item-03-pjm-study-results-additional-scenarios.ashx}.

\bibitem{DOEcp}
{Federal Energy Regulatory Commission, Department of Energy}.
\newblock Carbon pricing in organized wholesale electricity markets.
\newblock 2021.
\newblock \url{https://www.federalregister.gov/documents/2021/04/23/2021-08218/carbon-pricing-in-organized-wholesale-electricity-markets}.

\bibitem{scecp}
{Southern California Edison}.
\newblock Dynamic rate pilot.
\newblock 2021.
\newblock \url{https://www.dret-ca.com/dynamic-rate-pilot/}.

\end{thebibliography}
